\documentclass[journal]{IEEEtran}
\usepackage{algorithm}
\usepackage{algpseudocode}
\usepackage{tabularx}
\usepackage{graphicx}
\usepackage{amsfonts}
\usepackage{caption}
\usepackage{subcaption}
\usepackage{epstopdf}
\usepackage{breqn}
\usepackage{color,soul}
\usepackage{array}
\usepackage{authblk}
\usepackage{tabularx}
\usepackage{amsmath,graphicx}
\usepackage{mathtools}
\usepackage{amsthm}
\usepackage{breqn}
\usepackage{epstopdf}
\usepackage{array}
\usepackage{authblk}
\usepackage{mathtools} 
\usepackage{extarrows} 

\usepackage[justification=centering]{caption}

\usepackage{relsize}

\newcommand{\overbar}[1]{\mkern1.5mu\overline{\mkern-1.5mu#1\mkern-1.5mu}\mkern 1.5mu}

\begin{document}
%
\title{Time-Varying Graph Mode Decomposition}
%
%
%

\author{Naveed ur Rehman,~\IEEEmembership{Senior Member,~IEEE}
\thanks{N. Rehman is with the Department
of Electrical and Computer Engineering, Aarhus University, 8200 Aarhus N, Denmark e-mail: naveed.rehman@ece.au.dk.}}

\markboth{Submitted to IEEE Transactions on Signal Processing}%
{Naveed ur Rehman: Time-Varying Graph Mode Decomposition}
\maketitle

\begin{abstract}
Time-varying graph signals are alternative representation of multivariate (or multichannel) signals in which a single time-series is associated with each of the nodes or vertex of a graph. Aided by the graph-theoretic tools, time-varying graph models have the ability to capture the underlying structure of the data associated with multiple nodes of a graph -- a feat that is hard to accomplish using standard signal processing approaches. The aim of this contribution is to propose a method for the decomposition of time-varying graph signals into a set of \textit{graph modes}. The graph modes can be interpreted in terms of their temporal, spectral and topological characteristics. From the temporal (spectral) viewpoint, the graph modes represent the finite number of oscillatory signal components (output of multiple band-pass filters whose center frequencies and bandwidths are learned in a fully data-driven manner), similar in properties to those obtained from the empirical mode decomposition and related approaches. From the topological perspective, the graph modes quantify the functional connectivity of the graph vertices at multiple scales based on their signal content. In order to estimate the graph modes, a variational optimization formulation is designed that includes necessary temporal, spectral and topological requirements relevant to the graph modes. An efficient method to solve that problem is developed which is based on the alternating direction method of multipliers (ADMM) and the primal-dual optimization approach. Finally, the ability of the method to enable a joint analysis of the temporal and topological characteristics of time-varying graph signals, at multiple frequency bands/scales, is demonstrated on a series of synthetic and real time-varying graph data sets.

\end{abstract}


%
\IEEEpeerreviewmaketitle

\vspace{-3mm}
\section{Introduction}
\IEEEPARstart{G}raph signal processing (GSP) is an emerging multidisciplinary research area that deals with the concepts, theory and methods for studying signals residing on irregular structures or networks \cite{Ortega18}. Signals exhibiting those characteristics are now routinely collected from diverse sources and sensors in modern networks e.g., biological \cite{Huang18}, energy \cite{Zhao18}, 
sensor- and social-networks \cite{Ireneusz17}. 
Classical signal processing approaches, mainly designed for signals residing in the euclidean spaces (time-series and images), fall short of effectively modeling the signals defined on irregular structures and elucidating their complex interactions. 

To this end, GSP aims to extend classical signal processing tools to signals supported on irregular (graph) structures. This has led to new concepts and methods extending the harmonic analysis to graph signals e.g., graph Fourier transform \cite{Moura13}; filter bank \cite{Tremblay16}, compressed sensing and reconstruction of graph signals \cite{Weiyu11}; estimating graph spectrum through related new concepts of stationarity for graphs \cite{Perraudin17}; and vertex-frequency graph representations \cite{Stan21}. 

While the above concepts and methods are mainly applicable to \textit{static} graph signals i.e., those that do not evolve with time, recent efforts are also targeted towards the analysis of graph signals that are dynamic (or time-varying) in nature. Those are referred to as \textit{time-varying graph signals}. In essence, time-varying graph signals are intimately related to multivariate (or multichannel) signals that comprise time-series data associated with its each variate (or channel). Using graph-theoretic tools, such data could be alternatively modeled as time-varying graph signals in which each time-series is associated with a node (or a vertex) of a graph, with the topological interactions between the graph nodes being represented through an adjacency (or Laplacian) matrix. Indeed, the ability of graph-theoretic framework to effectively model the nodal connections within irregularly structured data makes it a more suitable candidate to handle multivariate signals as compared to the standard signal processing approaches. Some notable examples of graph-powered techniques for multivariate signals include frameworks for harmonic time-vertex analysis \cite{Grassi18}, time-varying graph reconstruction \cite{Giraldo22}, subsampling \cite{Jiang21} and denoising \cite{Tay21}. 
Still, graph-theoretic tools are largely an untapped resource when it comes to information extraction from multivariate signals.

Over the last couple of decades, one of the areas in non-stationary signal processing that has found interdisciplinary applications is the data-driven harmonic decomposition of non-stationary signals e.g., empirical mode decomposition (EMD) \cite{Huang98}, variational mode decomposition (VMD) \cite{Dragomiretskiy14}, synchrosqueezed transform (SST) \cite{DAUBECHIES11} and sliding-window singular spectrum analysis (SSA) \cite{Harmouche18}. Originally designed for single-channel time series, the overarching goal of signal decomposition methods is to decompose a signal into its weakly non-stationary and oscillatory amplitude- frequency-modulated (AM-FM) components in a fully data-driven manner i.e., without prior assumptions on input data. Recently, extensions of these methods for multivariate signals have also emerged, namely multivariate extensions of empirical mode decomposition (MEMD) \cite{Rehman10}, variational mode decomposition (MVMD) \cite{Rehman19} and iterative filtering \cite{Cicone22}; a recent article provides an overview of both univariate and multivariate signal decomposition methods along with their comparative analysis \cite{Rehman22}. As those are derived directly from their univariate counterparts that focused only on the temporal variations of a signal by design, the existing multivariate extensions fail to model complex structural (inter-channel) connections within data. As a result, none of the existing multivariate decomposition approaches are able to provide information about the rich functional connectivity patterns that maybe present in the data. In this paper, this issue is addressed through the unification of the concepts from GSP and harmonic analysis by modeling multivariate signals as time-varying graphs.

Specifically, this contribution proposes the first-ever model for the decomposition of time-varying graph signals into multiple \textit{graph modes} using variational optimization. The developed model includes information about the network (or topological) interactions as well as the dynamic (time-varying) behaviour of the data. The desired characteristics of the graph modes dictates the design of the resulting optimization formulation - reflecting both the temporal and spectral requirements as well as the topological constraints that the graph modes must exhibit. For instance, the temporal (and spectral) requirements include the separation of the input signal into a finite number of oscillatory (and band-limited) components that are derived in a fully data-driven manner i.e., without making any assumptions in the form of basis functions. Those components must also sum up to obtain the original signal (reconstruction). From the topological perspective, the following characteristics of the graph modes are reflected in the optimization formulation: i) the modes should include explicit information about the nodal connections within data, via associated adjacency or Laplacian matrix; ii) the modes should be \textit{smooth} over the respective graph structure. We employ the alternating direction method of multipliers (ADMM) and primal-dual approach to solve the resulting optimization formulation that alternates between the estimation of the band-limited signal components of the graph modes and the estimation of their respective topological structures. The resulting method, termed the time-varying graph mode decomposition (TVGMD), enables the joint analysis of signal's temporal dynamics and the corresponding network structures at multiple scales. 

\section{Related Work}
Recently, temporal graph signal decomposition (TGSD) method has been proposed for the decomposition of time-varying graph signals \cite{McNeil21}. The key idea behind this method is to encode an input signal through a combination of the fixed graph and fixed time dictionaries, resulting in a compact representation of multivariate data. In essence, the TGSD method cleverly extends the recent graph dictionary learning approaches to account for the decomposition of time-varying signals \cite{Elad19}. The proposed method (TVGMD), however, is fundamentally different and superior to the TGSD method in the following ways: i) TVGMD is fully data-driven, learning the inherent time-varying signal components as well as the associated connectivity graphs directly from the data, whereas the TGSD method uses over-complete dictionaries (or basis functions), making it a projection-based method; ii) TVGMD provides explicit information regarding the data connectivity of the graph nodes at multiple scales, which TGSD does not accomplish. 

A different class of multivariate signal decomposition methods are inspired from univariate data-driven signal decomposition approaches \cite{Rehman10,Rehman19,Cicone22}. To reiterate, this family of methods focus mainly on the temporal features of a signal, ignoring the rich topological structures that may exist between multiple data sources (channels). For instance, the only topological constraint included within the MVMD method is the enforcement of the same center frequency across all data channels, in a single output mode (or component) \cite{Rehman19}. Since the center frequency is a \textit{global} signal property, the MVMD (and other decomposition methods) fail to model, enforce or capture \textit{local} signal characteristics e.g., signal smoothness across connected data channels at multiple scales. Further, no direct mechanism exists to provide information regarding the functional connectivity of data by using these approaches.  

By modeling the multivariate signals as time-varying graph signals and designing a variational optimization formulation that includes both the temporal and topological constraints for \textit{graph modes}, the proposed TVGM method accomplishes a fully data-driven decomposition of time-varying graph signals. The method concurrently obtains multiple graph signal modes and the associated connectivity graphs which both the TGSD and multivariate decomposition methods fail to achieve.     

The rest of the paper is organized as follows: 
Section III reviews the preliminary concepts of graph signal processing (GSP), introduces the problem statement for TVGMD and specifies the associated optimization formulation. Section IV elucidates the steps to solve the TVGMD optimization problem. Section V demonstrates the utility of the TVGMD method through detailed experiments on a range of synthetic and real-life time-varying graph signals. Specifically, the real time-varying graph signals used in our experiments include electroencephalogram (EEG), electricity consumption signals and plant-wide oscillation data from an industrial setting. The paper concludes with the discussion and conclusion sections.          


\section{Time-varying Graph Mode Decomposition: Setting up the Optimization formulation}
\subsection{Preliminaries on Graph Signals}
We denote a weighted and undirected graph by $R=(\mathcal{V},\mathcal{E},W)$, where $\mathcal{V}$ represents the finite set of $N$ nodes or vertices and $\mathcal{E}\subset \mathcal{V}\times \mathcal{V}$ denotes the finite set of edges of the graph. $W\in \mathbb{R}^{N\times N}$ is the weighted and symmetric adjacency matrix of the graph. The entry $W_{mn}$ of the weighted adjacency matrix $W$ quantifies the weight of the edge between the $m$-th and the $n$-th vertices. These weights are generally non-negative and greater the weight, greater the connectivity (or similarity) between the corresponding vertices. $W_{mn}=0$ means that the vertices $m$ and $n$ are not connected.  

Another matrix that is widely used to represent the graph connectivity is the \textbf{graph Laplacian matrix}, denoted by $L$. It can be defined in terms of $W$ as $L=D-W$, where $D=$ \textbf{diag}$(d_1,d_2,\ldots,d_N)$ 
denotes the \textit{degree matrix} that is formed from the vertex degrees. Particularly, $d_n$, which is the $n$-th diagonal entry of $D$, is computed by taking the sum of weights of all the edges that are connected to the $n$-th vertex.    

We define a \textbf{graph signal} on $R$ as a function $u:\mathcal{V} \to \mathbb{R}$ which assigns a real (signal) value to each vertex. This way, a graph signal can be represented as a $N$-dimensional vector  $\mathbf{u}\in\mathbb{R}^N$ whose $n$-th entry, $u_n$, denotes the graph signal measurement (or value) at the $n$-th vertex.          

In a similar vein, a \textbf{time-varying graph signal} on $R$ can be defined as a function $\mathbf{u}:\mathcal{V} \to \mathbb{R}^T$ that assigns a (time-series) vector of length $T$, to each vertex. With a time-series assigned to each of the $N$ nodes of the graph, the resulting time-varying graph signal can be represented as a matrix $U\in\mathbb{R}^{N\times T}$. The $t$-th column of $U$, denoted by $\mathbf{u}_t$, represents the graph signal corresponding to the $t$-th time instant. Further, the $n$-th row of $U$, denoted by $\tilde{\mathbf{u}}_n$, represents the time-series corresponding to the $n$-th vertex of the graph $R$. In terms of notation for multivariate signal, $\tilde{\mathbf{u}}_n$ represents the $n$-th channel of the multivariate signal $U$. Note that in the limiting case of $T \to \infty$ and infinite sampling frequency, a continuous and unlimited-duration time-varying graph signal $\tilde{u}_n(t)$ will replace $\tilde{\mathbf{u}}_n$ - representing a continuous-time function rather than a vector, corresponding to the $n$-th vertex.

The \textbf{smoothness} of a time-varying graph signal $U$ on $R$ can be measured and quantified in terms of the following function of the graph Laplacian $L$

\begin{equation}
\begin{aligned}
Tr\big(U^T L U\big) &= \frac{1}{2}\sum_{m}\sum_{n}W_{mn}||\tilde{\mathbf{u}}_m-\tilde{\mathbf{u}}_n||^2\\
&= \frac{1}{2}||W \circ Z||_{1,1},
\label{eq:smooth}
\end{aligned}
\end{equation}

\noindent where $W_{mn}\in\mathbb{R}_+$ denotes the weight of the edge between the nodes $m$ and $n$. the resulting matrix $W\in\mathbb{R}^{N\times N}$ is the weighted graph adjacency matrix that is related to the graph Laplacian matrix via $L=D-W$. Further, $Z_{mn}=||\tilde{\mathbf{u}}_m-\tilde{\mathbf{u}}_n||^2$ are the elements of the pairwise distance matrix $Z\in\mathbb{R}_{+}^{N\times N}$, $\circ$ is the Hadamard product, $Tr(.)$ denotes the trace of a matrix and $||Z||_{1,1}$ is the element-wise norm-1 of $Z$. It can be verified that the smaller the value of the above measure \eqref{eq:smooth}, the greater the smoothness of the graph signal over $R$ i.e., strongly connected vertices (with higher corresponding edge weights) will have similar signal values.

\subsection{Signal Decomposition into Band-limited Modes}
Signal decomposition (SD) refers to the process of decomposing a uni- or multi-variate signal into its constituent finite number of band-limited modes or components. Let $\tilde{x}(t)$ and $\tilde{u}_k(t)$ denote the univariate input signal and its extracted $k$-th mode respectively, then SD results in the following

\begin{equation}
\tilde{x}(t)=\sum_{k=1}^{K}\tilde{u}^{(k)}(t).
\label{eq:vmd}
\end{equation}  

Several data-driven approaches to SD have emerged in recent years that accomplish this decomposition without making assumptions on data e.g., without basis functions or dictionaries. One of those methods is variational mode decomposition (VMD) \cite{Dragomiretskiy14} that uses a variational optimization program to obtain the desired modes. The cost function and the constraints of the optimization program respectively are the sum of the bandwidths of all modes $\mathlarger{\{}\tilde{u}^{(k)}(t)\mathlarger{\}}_{k=1}^K$ and the reconstruction of the signal from its modes. The optimization program for VMD is as follows   

\begin{equation}
\begin{aligned}
& \underset{\{\tilde{u}^{(k)}\},\{\omega^{(k)}\}}{\text{\normalsize{minimize}}}
& &\Bigg\{\mathlarger{\sum_k}\Bigg\Vert{\partial_t\Big[\tilde{u}_+^{(k)}(t)e^{-j\omega^{(k)}t}\Big]\Bigg\Vert^2_2}\Bigg\}\\[10pt]
& \text{subject to}
& & \sum_{k}\tilde{u}^{(k)}(t) = \tilde{x}(t),
\label{eq:vmd_opt}
\end{aligned}
\end{equation}

\noindent where $u_+^{(k)}(t)$ denotes an analytic signal corresponding to $u^{(k)}(t)$; the symbol $\partial_t$ represents partial derivative operation with respect to time; $\{u^{(k)}\}$ and $\{\omega^{(k)}\}$ respectively denote sets of all $K$ number of modes and their center frequencies. The squared $l^2$-norm term within the parenthesis in the cost function \eqref{eq:vmd_opt} is an estimator of the sum of the bandwidths of all the signal modes.

An extension of VMD to multivariate signals was developed in \cite{Rehman19} that utilized a model for multivariate modulated oscillations to design the following optimization formulation for the decomposition of multivariate signals. 

\begin{equation}
\begin{aligned}
& \underset{\{u_n^{(k)}\},\{\omega^{(k)}\}}{\text{minimize}}
& & \Bigg\{ \mathlarger{\sum_{k}\sum_n}\Bigg\Vert\partial_t\Big[\tilde{u}_{n+}^{(k)}(t) e^{-j\omega^{(k)}t}\Big]\Bigg\Vert^2_2\Bigg\}\\[10pt]
& \text{subject to}
& & \sum_{k}\tilde{u}_n^{(k)}(t) = \tilde{x}_n(t),\text{   } n=1,2, \ldots ,N.
\label{eq:mvmd_opt}
\end{aligned}
\end{equation}

Note that the main focus of MVMD is on extracting the temporal features of a signal i.e., band-limited oscillations. The only topological constraint that it enforces on a signal mode is to have the same fixed center frequency $\omega^{(k)}$ across all its channels. Clearly, that is inadequate to model and extract rich topological structures that time-varying graph signals typically possess.        

\subsection{Problem Statement}
Consider a time-varying graph signal $\mathbf{\tilde{x}(t)}$ that comprises a set of time-series signals: $\mathbf{\tilde{x}(t)}=[\tilde{x}_1(t), \ldots, \tilde{x}_N(t)]$, with $\tilde{x}_n(t)$ representing the time-series corresponding to the $n$-th vertex. In the discrete domain, $\mathbf{\tilde{x}(t)}$ could also be represented by a matrix $\mathbf{X}\in\mathbb{R}^{N\times T}$ where $T$ denotes the number of time samples of the time-varying graph signal. 

Let the graph signal $\mathbf{\tilde{x}(t)}$ be decomposed into a set of band-limited oscillatory components as follows

\begin{equation}
\mathbf{\tilde{x}}(t)=\sum_{k=1}^{K}\mathbf{g}^{(k)}(t),
\label{eq:mvmd}
\end{equation} 

\noindent where $\mathbf{g}^{(k)}(t)=[g_1^{(k)}(t),g_2^{(k)}(t),\ldots,g_N^{(k)}(t)]$. The set of \textit{graph modes} of $\mathbf{\tilde{x}(t)}$ comprise its band-limited constituent components along with their weighted adjacency or \textit{connectivity} matrices. For instance, the $k$-th graph mode is represented by $G^{(k)}=[\mathbf{g}^{(k)}(t), W^{(k)}]$, with the collection of all the graph modes of $\mathbf{\tilde{x}(t)}$ be denoted by $\mathbf{G}=[G^{(1)}, G^{(2)}, \ldots, G^{(K)}]$. \textit{Concretely, given the number of modes $K$ as defined by the user, the problem here is to decompose a time-varying graph signal $\mathbf{\tilde{x}(t)}$ into its constituent graph modes} 
\begin{equation}
\mathbf{\tilde{x}(t)}\to[G^{(1)}, G^{(2)}, \ldots, G^{(K)}].
\end{equation}

\subsection{Optimization Formulation for Time-Varying Graph Mode Decomposition (TVGMD)}
To obtain the desired graph modes $\mathbf{G}$ from input time-varying graph signal $\mathbf{\tilde{x}(t)}$, we design a variational optimization program that includes both the temporal and the topological constraints for the extraction of physically meaningful modes. Specifically, the temporal requirements include the extraction of AM-FM oscillatory components, $\mathbf{g}^{(k)}(t)=[g_1^{(k)}(t),g_2^{(k)}(t),\ldots,g_N^{(k)}(t)]$, which are band-limited in nature, with the center frequency and the bandwidth of each mode being estimated via a data-driven approach. Another temporal requirement is the exact reconstruction of the original graph signal from its extracted modes. 

Next, topological or structural constraints related to the time-varying graph modes have been included in the optimization formulation. Particularly, in the proposed optimization program, we embed the requirement of the \textit{smoothness} of each extracted mode $\mathbf{g}^{(k)}(t)$ with respect to its graph structure $W^{(k)}$. To this end, the (non-)smoothness measure \eqref{eq:smooth} corresponding to the $K$ graph modes $\{G^{(k)}\}_{k=1}^K$ is included in the optimization program. Finally, the optimization program includes 
requirements for the validity of the graph adjacency matrices. The resulting optimization formulation for time-varying graph mode decomposition (TVGMD) method is given as follows

\begin{align}
\begin{split}
\underset{\{\tilde{g}_n^{(k)}\},\{\omega^{(k)}\},\{W^{(k)}\}}{\text{minimize}}\quad &
\Bigg\{\alpha \mathlarger{\sum_{k=1}^K\sum_{n=1}^N}\Big\Vert\partial_t\Big[\tilde{g}_{n+}^{(k)}(t) e^{-j\omega^{(k)}t}\Big]\Big\Vert^2_2\Bigg\}\\
&\quad + \beta\mathlarger{\sum_{k=1}^K}||W^{(k)} \circ Z^{(k)}||_{1,1}\\
&\quad +\frac{\gamma}{2}\mathlarger{\sum_{k=1}^K}\lVert W^{(k)}\rVert^2_F-\mathlarger{\sum_{k=1}^K}\mathbf{1}^T\log(W^{(k)}\mathbf{1})
\end{split}
\label{cost1} 
\\[2ex]
\text{subject to} \quad &
\sum_{k}\tilde{g}_n^{(k)}(t) = \tilde{x}_n(t),\quad \forall n  \\ 
& W^{(k)}\geq\mathbf{0}, \quad W^{(k)}=W^{(k)^T}, \notag\\  & \text{diag}(W^{(k)})=0, \quad \forall k \notag
\end{align}

\noindent where $\tilde{g}_{n+}^{(k)}(t)$ denotes an analytical signal corresponding to $\tilde{g}_{n}^{(k)}(t)$. The output of the optimization program includes the graph modes $\{G^{(k)}\}_{k=1}^K$ that comprise the time-varying oscillatory graph components, $\{\mathbf{g}^{(k)}\}_{k=1}^K$, along with their respective connectivity structures $\{W^{(k)}\}_{k=1}^K$. The center frequency $\omega^{(k)}$ of each of the extracted graph modes is also estimated. 

Some comments on the TVGMD optimization formulation are in order. The first term of the cost function \eqref{cost1} represents the sum of the the bandwidths of all the extracted graph modes that ensures the extraction of band-limited oscillatory components in the temporal domain \cite{Rehman19}. The second term in the cost function imposes the smoothness of the extracted graph modes over their respective graph structures $W^{(k)}$. This way, valid graph structures $W^{(k)}$ are obtained i.e., for a pair of nodes with similar signal values, the corresponding adjacency matrix element has a higher weight. Further, the term involving the Frobenius norm of $W^{(k)}$ prohibits the formation of very large edge weights in $W^{(k)}$ while not penalizing the smaller weights. The last term can be interpreted as a logarithmic barrier on the node degree vector, enforcing the positive degrees on the graph nodes while not preventing the individual connections from becoming zero; this results in a better connected graph \cite{kalofolias16}. 

As for the constraints of the optimization problem, the first constraint guarantees that the obtained graph modes fully reconstruct the original graph signal. The remaining constraints relate to the validity of the adjacency matrices at multiple scales, including their symmetricity, non-negative values and a zero vector at the diagonal.  

\section{Optimization Solution}
The first step towards the solution of the TVGMD optimization problem is to convert the original constrained optimization formulation to the unconstrained formulation. Then, a combination of the alternating direction method of multipliers (ADMM) and the primal-dual approach is used to solve the problem. 

Note that in the original optimization problem there are constraints related to the symmetricity of $W^{(k)}$ along with a zero vector as its diagonal. Those constraints are cumbersome while searching for a convenient solution to the problem. To address this problem, it was proposed in \cite{kalofolias16} to use a vector representation of $W^{(k)}$  within the space $\mathcal{W}_v=w\in\mathbb{R}_+^{N(N-1)/2}$. Following that approach, let $w^{(k)}$ and $z^{(k)}$ denote the vectors comprising all entries above the main diagonal of $W^{(k)}$ and $Z^{(k)}$ respectively. Further, $Q^{(k)}$ denotes a binary matrix that fulfills the condition $Q^{(k)}w^{(k)}=W^{(k)}\mathbf{1}$. This way, $Q^{(k)}w^{(k)}\in\mathbb{R}^N$ has a convenient interpretation as a vector of node degrees for $\mathbf{g^{(k)}}(t)$. Then, the original optimization problem can be modified as follows

\begin{align}
\begin{split}
\underset{\{\tilde{g}_n^{(k)}\},\{\omega^{(k)}\},\{w^{(k)}\}}{\text{minimize}}\quad &
\Bigg\{\alpha \mathlarger{\sum_{k=1}^K\sum_{n=1}^N}\Big\Vert\partial_t\Big[\tilde{g}_{n+}^{(k)}(t) e^{-j\omega^{(k)}t}\Big]\Big\Vert^2_2\Big\}\\
&\quad + \beta\mathlarger{\sum_{k=1}^K} 2w^{(k)T}z^{(k)}\\
&\quad +\gamma\mathlarger{\sum_{k=1}^K}\lVert w^{(k)}\rVert^2_2-\mathlarger{\sum_{k=1}^K}\mathbf{1}^T\log(Q^{(k)}w^{(k)})
\end{split}
\label{cost2} 
\\[2ex]
\text{subject to} \quad &
\sum_{k}\tilde{g}_n^{(k)}(t) = \tilde{x}_n(t),\quad \forall n  \\ 
& w^{(k)}\geq\mathbf{0}. \quad \forall k  \notag
\end{align}


By introducing the sets of dual variables $\lambda_n(t)$, corresponding to each of the equality constraints in (10), the augmented Lagrangian function of the above optimization formulation can be written as 

\begin{equation}
\begin{aligned}
\mathcal{L}(\{\tilde{g}_n^{(k)}\},\{\omega^{(k)}\},\{w^{(k)}\}) = h_1(\{\tilde{g}_n^{(k)}\},\{\omega^{(k)}\})+ h_2(\{w^{(k)}\})
\label{eq:tgmd-lag}
\end{aligned}
\end{equation}
where 

\begin{dmath}
h_1(\{\tilde{g}_n^{(k)}\},\{\omega^{(k)}\})=\alpha \mathlarger{\sum_{k=1}^K\sum_{n=1}^N}\Big\Vert\partial_t\Big[\tilde{g}_{n+}^{(k)}(t) e^{-j\omega^{(k)}t}\Big]\Big\Vert^2_2
+\mathlarger{\sum_n}\Big\Vert \tilde{x}_n(t)-\sum_{k}\tilde{g}_{n}^{(k)}(t)\Big\Vert_2^2+ \mathlarger{\sum_n}\Big\langle\lambda_n(t),\tilde{x}_n(t)-\sum_{k}\tilde{g}_{n}^{(k)}(t)\Big\rangle,
\label{eq:g1}
\end{dmath}
\begin{dmath}
h_2(\{w^{(k)}\})=\beta\mathlarger{\sum_{k=1}^K} 2w^{(k)T}z^{(k)}+\gamma\mathlarger{\sum_{k=1}^K}\lVert w^{(k)}\rVert^2_2-\mathlarger{\sum_{k=1}^K}\mathbf{1}^T\log(Q^{(k)}w^{(k)})+\mathlarger{\sum_{k=1}^K}\mathbb{I}_{\{w^{(k)}\geq0\}}
\label{eq:g2}
\end{dmath}

\noindent with $\mathbb{I}_{\{w\geq0\}}$ representing the indicator function that is equal to zero for $w\geq0$ but is otherwise infinite. The term involving the indicator function has been added to cater for the inequality constraints in (10). Note the difference between the symbols $\omega^{(k)}$ and $w^{(k)}$ that respectively denote the center frequency and the weight vector (corresponding to $W^{(k)}$) of the $k$-th graph mode.   

\begin{algorithm}[t]
\caption{\bf ADMM Optimization Strategy for TVGMD}
\label{alg:tvgmd1}
\vspace{2mm}\text{Initialize: } \vspace{-2.5mm}\begin{equation*}\{\tilde{g}_{n,1}^{(k)}\}\gets 0\mbox{, }\{\omega^{(k)}_1\}\gets 0\mbox{, }\{w^{(k)}_1\}\gets 0\mbox{, }\lambda_{n,1}\mbox{, }i\gets 0\mbox{, } \epsilon\gets 10^{-7}\end{equation*}
\vspace{-3mm}
\begin{algorithmic} 
	\Repeat
	\State $i \gets i+1$
	\vspace{.2cm}
	\For{$k=1:K$}
	\vspace{.1cm}
	\For{$n=1:N$}
	\vspace{.1cm}
	\textit{Update graph modes in time $\tilde{g}_{n}^{(k)}$:}
	\State  \begin{equation} \tilde{g}_{n,i+1}^{(k)}/\overbar{U}_{i+1}^{(k)}\gets\underset{\tilde{g}_{n}^{(k)}}{\text{arg min }} \mathcal{L}\left(\mathlarger{\{}\tilde{g}_{n,i}^{(k')}\mathlarger{\}},\mathlarger{\{}\omega_{i}^{(k')}\mathlarger{\}},\mathlarger{\{}w_{i}^{(k')}\mathlarger{\}},\lambda_{n,i}\right) \label{eq:tgmd_uk}  \end{equation}
	\EndFor
	\vspace{.1cm}
	\EndFor
	\vspace{.1cm}
	\For{$k=1:K$}
	\vspace{.1cm}
	\textit{Update center frequency $\omega^{(k)}$:}
	\State \begin{equation}\omega_{i+1}^{(k)}\gets\underset{\omega^{(k)}}{\text{arg min }} \mathcal{L}\left({\{\tilde{g}_{n,i+1}^{(k')}\},\{\omega_{i}^{(k')}\},\{w_{i}^{(k')}\},\lambda_{n,i}}\right) \label{eq:tgmd_freqtime} \end{equation} 
	\EndFor
	\vspace{.2cm}
	\For{$k=1:K$}
	\vspace{.1cm}
	\textit{Update modes along graph geodesics:}
	\State \begin{equation} U_{i+1}^{(k)}\gets\underset{\overbar{U}^{(k)}}{\text{arg min }} \mathcal{L}\left(\mathlarger{\{}\overbar{U}_{i+1}^{(k')}\mathlarger{\}},\mathlarger{\{}\omega_{i+1}^{(k')}\mathlarger{\}},\mathlarger{\{}w_{i}^{(k')}\mathlarger{\}},\lambda_{n,i}\right) \label{eq:tgmd_geodesicstime}  \end{equation} 
	\EndFor
	\vspace{.1cm}
	\vspace{.1cm}
	\For{$k=1:K$}
	\vspace{.1cm}
	\textit{Update graph adjacency matrix $w^{(k)}$:}
	\State \begin{equation}w_{i+1}^{(k)}\gets\underset{w^{(k)}}{\text{arg min }} \mathcal{L}\left({\{\tilde{g}_{n,i+1}^{(k')}\},\{\omega_{i+1}^{(k')}\},\{w_{i}^{(k')}\},\lambda_{n,i}}\right) \label{eq:tgmd_wtime} \end{equation} 
	\EndFor

	\vspace{.1cm}
	\For{$n=1:N$}
	\vspace{.1cm}
	\textit{Update $\lambda_n$:}
	\State \begin{equation}\lambda_{n,i+1}=\lambda_{n,i}+\tau \Big(\tilde{x}_n-\sum_k \tilde{g}_{n,i+1}^{(k)}\Big) \label{eq:tgmg_lambdatime} \end{equation} 	
	\EndFor
	\vspace{.1cm}
	\Until{Convergence: $\sum_k\sum_n \frac{\Vert \tilde{g}_{n,i+1}^{(k)}-\tilde{g}_{n,i}^{(k)}\Vert_2^2}{\Vert \tilde{g}_{n,i}^{(k)}\Vert_2^2}<\epsilon$} 
\end{algorithmic}
\end{algorithm}

The unconstrained optimization problem \eqref{eq:tgmd-lag} is solved using the alternating direction method of multipliers (ADMM) \cite{Boyd11} and the primal-dual approach \cite{Komodakis15}. The ADMM operates by converting a complex optimization problem into a series of simpler sub-optimization problems. Our approach towards the solution of \eqref{eq:tgmd-lag} is outlined in the Algorithm 1. By using the ADMM approach, the TVGMD unconstrained optimization problem \eqref{eq:tgmd-lag} is transformed into 4 simple optimization problems, relating to the: i) update of the graph modes $\tilde{g}_{n}^{(k)}$ in the time domain \eqref{eq:tgmd_uk}; ii) update of the center frequency $\omega^{(k)}$ for each graph mode \eqref{eq:tgmd_freqtime}; iii) update of the graph modes along the graph geodesics or vertices $U^{(k)}$ \eqref{eq:tgmd_geodesicstime}; and iv) update of the adjacency or Laplacian matrix $W^{(k)}$ for each graph mode \eqref{eq:tgmd_wtime}. Finally, the dual variables $\lambda_{n}$ are updated according to \eqref{eq:tgmg_lambdatime}. In the following, we obtain the solutions to the above optimization sub-problems.

\subsection{Graph mode update in the time domain:}  
To solve the minimization problem related to the graph mode update in the time domain \eqref{eq:tgmd_uk}, we write the associated optimization formulation below 
    
\begin{equation}
\begin{aligned}
\tilde{g}_{n,i+1}^{(k)}=\underset{\tilde{g}_{n}^{(k)}}{\text{arg min}} \Bigg\{\alpha\Big\Vert\partial_t\Big[\tilde{g}_{n+}^{(k)}(t) e^{-j\omega^{(k)}t}\Big]\Big\Vert^2_2+\\
\Big\Vert \tilde{x}_n(t)-\mathlarger{\sum_{k'}} \tilde{g}_{n}^{(k')}(t)+\frac{\lambda_n(t)}{2}\Big\Vert^2_2\Bigg\}.
\end{aligned}
\end{equation}

The above problem was encountered within the MVMD optimization formulation \cite{Rehman19} and was solved in the Fourier (or frequency) domain. Using that result, the following graph mode update relation (in the time domain) is obtained

\begin{equation}
\begin{aligned}
\hat{g}_{n,i+1}^{(k)}(\omega) = \frac{\hat{x}_n(\omega)-\sum_{k'\neq k}\hat{g}_{n}^{(k')}(\omega)+\frac{\hat{\lambda}_n(\omega)}{2}}{1+2\alpha(\omega-\omega^{(k)})^2},
\end{aligned}
\label{eq:tgmd_modeupdate}
\end{equation}

\noindent where $\hat{g}_{n}^{(k)}(\omega)$ denotes the Fourier transform of the $k$-th graph mode, $\tilde{g}_{n}^{(k)}(t)$.  

\subsection{Center frequency update:}
The center frequency of each graph mode is updated by solving the optimization problem in  \eqref{eq:tgmd_freqtime}. Here, only the first term of \eqref{eq:g1} depends on $\omega_k$, simplifying \eqref{eq:tgmd_freqtime} to

\begin{equation}
\omega_{n,i+1}^{(k)}=\underset{\omega^{(k)}}{\text{arg min}}\Big\{\mathlarger{\sum_n}\Big\Vert\partial_t\Big[\tilde{g}_{n+}^{(k)}(t) e^{-j\omega^{(k)}t}\Big]\Big\Vert^2_2\Bigg\}.
\label{eq:tgmd_wkupdate}
\end{equation}

Like the mode update problem, the above optimization problem was also solved in the frequency domain within the MVMD optimization solution; see (30)-(32) in \cite{Rehman19}. The resulting update relation for the center frequency $\omega^{(k)}$ of the $k$-th graph mode is given by  

\begin{equation}
\omega_{i+1}^{(k)}=\frac{\mathlarger{\sum_n}\mathlarger\int^{\infty}_0 \omega\times|\hat{g}_{n}^{(k)}(\omega)|^2 d\omega}{\mathlarger{\sum_n}\mathlarger\int^{\infty}_0 |\hat{g}_{n}^{(k)}(\omega)|^2 d\omega}.
\label{eq:mvmd_frequpdate}
\end{equation}

\subsection{Graph mode update along the graph geodesics:}
The next step involves the graph mode update along the graph geodesics (or vertices). This is fundamentally different from the mode update in the time domain which was conveniently performed through the Fourier transformation via \eqref{eq:tgmd_modeupdate}. In the time domain, the key requirement was the small bandwidth of each decomposed graph mode which was imposed through the first term of \eqref{eq:g1}. 

On the contrary, the crucial constraint for the graph modes update along the graph geodesics is that the obtained modes should be consistent with the associated connectivity matrices (e.g., adjacency or the Laplacian matrix). Simply put, the graph modes (along the graph geodesics) must be updated in such a way that they have similar values on the pair of nodes having a strong connection. This key requirement is implicitly encoded in the second terms of the TVGMD cost functions in \eqref{cost1} and \eqref{cost2}, and the first term of \eqref{eq:g2} that is part of the augmented Lagrangian function \eqref{eq:tgmd-lag}. In those terms, $Z^{(k)}$ and its corresponding vector $z^{(k)}$ are directly related to the graph modes $\tilde{g}^{(k)}(t)$ via \eqref{eq:smooth}. That is, $Z^{(k)}$ is the pairwise distance matrix of the graph modes at different vertices.    

Based on those observations, the corresponding optimization problem for the mode update along graph geodesics can be written as

\begin{equation}
\tilde{u}_{n,i+1}^{(k)}=\underset{\tilde{g}_{n}^{(k)}}{\text{arg min}}\Bigg\{\mathlarger{\sum_n}\Big\Vert \tilde{x}_n(t)-\sum_{k'}\tilde{g}_{n}^{(k')}(t)\Big\Vert_2^2 + 2\beta w^{(k)T}z^{(k)}\Bigg\}.
\label{eq:tgmd_geo1}
\end{equation}

If we let $\overbar{U}_{i+1}^{(k)}\}_{k=1}^K$ denote the matrix representation of the graph modes obtained by solving \eqref{eq:tgmd_uk}, and further by using \eqref{eq:smooth}, the above optimization problem can be expressed as

\begin{dmath}
U_{i+1}^{(k)}=\underset{\overbar{U}^{(k)}}{\text{arg min}}\Bigg\{\Big\Vert X-\mathlarger{\sum_{k'}}\overbar{U}_{i+1}^{(k')}\Big\Vert_2^2+2\beta Tr\left[\overbar{U}_{i+1}^{(k)T}L_i^{(k)}\overbar{U}_{i+1}^{(k)}\right]\Bigg\}.
\label{eq:tgmd_geo2}
\end{dmath}

By denoting $F^{(k)}=X-\sum_{k'\neq k}\overbar{U}_{i+1}^{(k')}$, the first term on the right hand side of \eqref{eq:tgmd_geo2} can be written as $\Big\Vert F^{(k)}-\overbar{U}_{i+1}^{(k)}\Big\Vert_2^2=Tr\left[(F^{(k)}-\overbar{U}_{i+1}^{(k)})(F^{(k)}-\overbar{U}_{i+1}^{(k)})^T\right]$, leading to the following optimization problem

\begin{dmath}
U_{i+1}^{(k)}=\underset{\overbar{U}^{(k)}}{\text{arg min}}\Big\{Tr\left[\Big(F^{(k)}-\overbar{U}_{i+1}^{(k)}\Big)\Big(F^{(k)}-\overbar{U}_{i+1}^{(k)}\Big)^T\right]+2\beta Tr\left[\overbar{U}_{i+1}^{(k)T}L_i^{(k)}\overbar{U}_{i+1}^{(k)}\right]\Big\}.
\label{eq:tgmd_geo3}
\end{dmath}

To minimize the above function that is clearly convex, we set its derivative to zero resulting in the following update for the graph modes along the graph geodesics

\begin{dmath}
U_{i+1}^{(k)}=\left(I+\beta L_i^{(k)}\right)^{-1}F^{(k)},
\label{eq:tgmd_geo3}
\end{dmath}

\noindent where $I$ denotes the identity matrix.

\subsection{Graph adjacency (connectivity) matrix update:}
The connectivity matrices of the graph modes are updated and obtained by solving the optimization (sub-)problem given by \eqref{eq:tgmd_wtime}. For our unconstrained Lagrangian function \eqref{eq:tgmd-lag}, the optimization problem \eqref{eq:tgmd_wtime} becomes

\begin{dmath}
w_{i+1}^{(k)}=\underset{w^{(k)}}{\text{arg min }} 2\beta w^{(k)T}z^{(k)}+\gamma\lVert w^{(k)}\rVert^2_2-\mathbf{1}^T\log(Q^{(k)}w^{(k)})+\mathbb{I}_{\{w^{(k)}\geq0\}}.
    \label{eq:connect1}
\end{dmath}

This problem is similar in form to the popular graph learning model proposed by Kalofolias in \cite{kalofolias16}. The only difference between the two models is the appearance of the parameter $\beta$ in the first term of \eqref{eq:connect1}, which was effectively equal to 1 in Kalofolias' model. That problem is solved by using the primal-dual optimization technique, illustrated in Algorithm 6 in \cite{Komodakis15}. We use the same technique to solve \eqref{eq:connect1}. Particularly, to make our problem amenable to be solved by using the primal-dual optimization technique, we cast our optimization problem as a sum of the following three functions

\begin{dmath}
    w_{i+1}^{(k)}=\underset{w^{(k)}}{\text{arg min }}b_1(w)+b_2(Kw)+b_3(w),
    \label{eq:primal-dual}
\end{dmath}
where
\begin{align}
b_1(w^{(k)})&= \mathbb{I}_{\{w^{(k)}\geq0\}}+2\beta w^{(k)T}z^{(k)},\notag\\
b_2(Q^{(k)}w^{(k)})&= -\mathbf{1}^T\log(Q^{(k)}w^{(k)}),\notag\\
b_3(w^{(k)})&=\gamma\lVert w^{(k)}\rVert^2.\notag
\end{align}

Using the above form of the optimization problem, the primal-dual algorithm used for the solving that problem is given in the Appendix 1.  

With the knowledge of the solutions to all the sub-optimization problems (14)-(17) within the TVGMD algorithm, the steps to apply the TVGMD algorithm are listed in Algorithm 2. In the algorithm, note that the TVGMD updates the graph modes in two steps: the first update \eqref{eq:tgmd_hatuk} relates to the requirements of the modes in the time domain, e.g., mainly oscillatory modes having small bandwidths, and is conducted in the Fourier domain. The second update \eqref{eq:tgmd_geodesic} concerns the mode constraints along the graph geodesics and is performed in the spatial domain. Between the two updates, a conversion step \eqref{eq:f2t} is needed that transforms the modes from the frequency to the time domain in each iteration.

\begin{algorithm}[H]
	\caption{\bf Time-varying Graph Mode Decomposition}
	\label{alg:tgmd}
	\vspace{2mm}\text{Input and user-defined parameters: $K, \alpha, \beta, \gamma\mbox{,}$} \vspace{-1.5mm}\begin{equation*}X=\{x_{n}(t)\}_{n=1}^N\xlongleftrightarrow{FT}\{\hat{x}_{n}(\omega)\}_{n=1}^N\end{equation*}
	
	\text{Initialize: } \vspace{-1.0mm}\begin{equation*}\{\hat{g}_{n,1}^{(k)}\}\gets 0\mbox{, }\{\omega^{(k)}_1\}\gets 0\mbox{, }\{W^{(k)}_1\}\gets 0\mbox{, }\lambda_{n,1}\mbox{, }i\gets 0\mbox{, } \epsilon\gets 10^{-7}\end{equation*}
	\vspace{-3mm}
	
	\begin{algorithmic} 
		\Repeat 
		\State $i \gets i+1$
		\For{$k=1:K$}
		\For{$n=1:N$}
		\textit{Updating graph mode $\hat{g}_{n}^{(k)}$:}
		\State  
		\begin{equation}  
		    \hat{g}_{n,i+1}^{(k)}(\omega) \gets \frac{\hat{x}_n(\omega)-\sum_{k'\neq k}\hat{g}_{n}^{(k')}(\omega)+\frac{\hat{\lambda}_n(\omega)}{2}}{1+2\alpha(\omega-\omega^{(k)})^2} 
		\label{eq:tgmd_hatuk}  
		\end{equation} 
		\EndFor
		\EndFor
		\vspace{.2cm}
		\For{$k=1:K$}
		\textit{Updating center frequency $\omega_{k}$:}
		\State 
		\begin{equation} 
		\omega_{n,i+1}^{(k)}\gets \frac{\mathlarger{\sum_n}\mathlarger\int^{\infty}_0 \omega|\hat{g}_{n,i}^{(k)}(\omega)|^2 d\omega}{\mathlarger{\sum_n}\mathlarger\int^{\infty}_0 |\hat{g}_{n,i}^{(k)}(\omega)|^2 d\omega}\notag
		\label{eq:tgmd_hatwk} 
		\end{equation} 
		\vspace{-0.4cm}
		\EndFor
		
		\State \begin{equation} 
		\Big\{\overbar{U}^{(k)}\Big\}_{k=1}^K\xlongleftarrow{IFT}\Big\{\hat{g}_{n,i+1}^{(k)}(\omega)\Big\}_{k=1}^K 
		\label{eq:f2t}
		\end{equation}
		\vspace{.1cm}
		\For{$k=1:K$}
	    \textit{Updating modes along geodesics:}
        \vspace{-0.3cm}
	    \State \begin{eqnarray}
	    L_i^{(k)}&\gets&D_i^{(k)}-W_i^{(k)} \notag\\
	    U_{i+1}^{(k)}&\gets&\left(I+\beta L_i^{(k)}\right)^{-1}\Big\{X-\sum_{k'\neq k}\overbar{U}^{(k')}\Big\} 
	    \label{eq:tgmd_geodesic} \end{eqnarray} 
	    \vspace{-0.3cm}
	    \EndFor
	    
	    \vspace{.2cm}
		\For{$k=1:K$}
	    \textit{Updating adjacency matrix:}
		\vspace{-0.4cm}
	    \State \begin{eqnarray}
	    W_i^{(k)}&\gets& L_i^{(k)} \notag\\
	    W_{i+1}^{(k)}&\gets&\mbox{Apply Algorithm 3 in the Appendix} \notag
            \end{eqnarray} 
	    \EndFor
		
		\State \begin{equation}
		\Big\{\hat{g}_{n,i+1}^{(k)}(\omega)\Big\}_{k=1}^K\xlongleftarrow{FT}\Big\{U_{i+1}^{(k)}\Big\}_{k=1}^K \notag
		\end{equation} 
		
		\vspace{.1cm}
		\For{$n=1:N$}
		\textit{Update $\hat{\lambda}_c$}:
		\vspace{-0.3cm}
		\State \begin{equation}\hat{\lambda}_{n,i+1}(\omega)=\hat{\lambda}_{n,i}(\omega)+\tau \Big(\hat{x}_n(\omega)-\sum_k \hat{g}_{n,i+1}^{(k)}(\omega)\Big) \notag \label{eq:tgmd_lambda} \end{equation} 
		\vspace{-0.3cm}
		\EndFor
		\Until{Convergence: $\sum_k\sum_n \frac{\Vert \hat{g}_{n,i+1}^{(k)}-\hat{g}_{n,i}^{(k)}\Vert_2^2}{\Vert \hat{g}_{n,i+1}^{(k)}\Vert_2^2}<\epsilon$} 
\end{algorithmic}
\end{algorithm}

Note that the TVGMD optimization scheme is not strictly ADMM owing to the: i) non-convexity of the original optimization cost function \eqref{cost2}; ii) deviation of our optimization formulation from the standard ADMM formulation that requires multiple set of variables having separable objective functions. Therefore, the algorithm is not guaranteed to converge to the global minimum. That said, we take inspiration from the fact that the ADMM approach has been successfully employed in several popular heuristics algorithms in signal processing that use non-convex optimization, including those based on signal decomposition \cite{Dragomiretskiy14} and graph learning \cite{Elad19}. 

Finally, some comments on the choice of the parameters of the TVGMD algorithm are in order. The $\alpha$ parameter determines the bandwidth of the decomposed graph modes; the smaller the value of $\alpha$, the higher the bandwidth of the extracted modes. For closely spaced graph modes in the frequency domain, choosing a higher value of $\alpha$ is recommended. The $\beta$ parameter dictates the smoothness of the graph modes $\mathbf{g}^{(k)}(t)$ with respect to their adjacency matrices $W^{(k)}$ -- an important parameter to ensure that the adjacency (connectivity) matrices at each scale are valid. As for the $\gamma$ parameter, it controls the sparsity of the obtained adjacency matrices: $\gamma=0$ leads to very sparse connectivity graphs whereas higher values tend to produce dense edge patterns in the graphs. The parameter $\tau$ offers a trade off between the exact reconstruction of the input graph signal (via the summation of its graph modes) and denoising. In the presence of noise in the input signal, choosing a very small value of $\tau$ (or even $\tau=0$) may be appropriate since exactly reconstructing a noisy input signal will not be desirable. In section VI, further suggestions regarding the choice of parameters in the TVGMD method are given.

\section{Experiments, Results and Applications}
\label{Sec4}
This section presents the results of experiments conducted on synthetic and real time-varying graph signals to demonstrate the power and the performance of the TVGMD algorithm. To our knowledge, the TVGMD algorithm is the first to concurrently obtain the constituent oscillatory components of a time-varying graph signal and their corresponding connectivity structures in the spatial domain, via a joint optimization scheme. To this end, the designed experiments and the chosen applications will highlight the ability of the TVGMD algorithm to extract both the time-domain and the graph-related (spatial) features from a range of input data. Those signals include the synthetic time-varying oscillatory graph signals, resting-state electroencephalogram (EEG) signals, electrical consumption data of clients within a power network and plant-wide oscillations from industrial networks. In Section VI, we present results that compare the proposed method with the only available temporal graph signal decomposition algorithm (TGSD) method \cite{McNeil21} and highlight their crucial differences. 
\begin{figure}[h]
\includegraphics[width=0.25\textwidth]{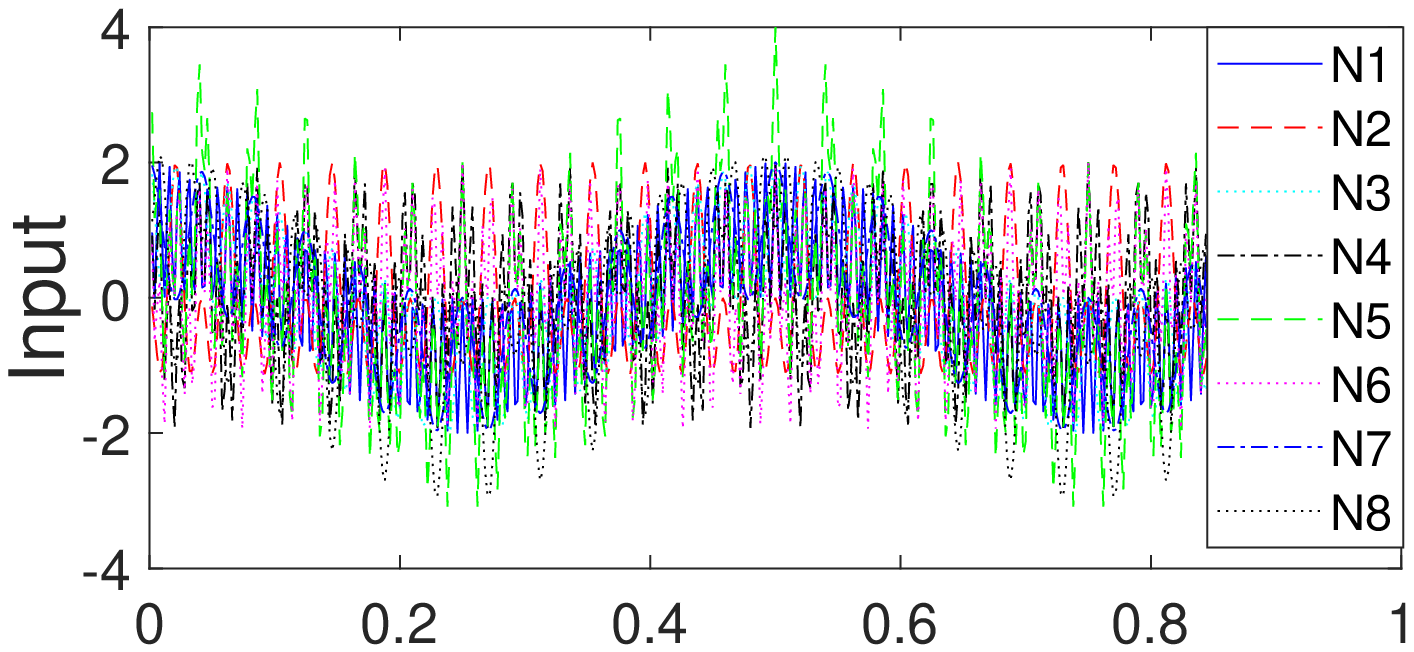}\hspace{-4mm}
\includegraphics[width=0.25\textwidth]{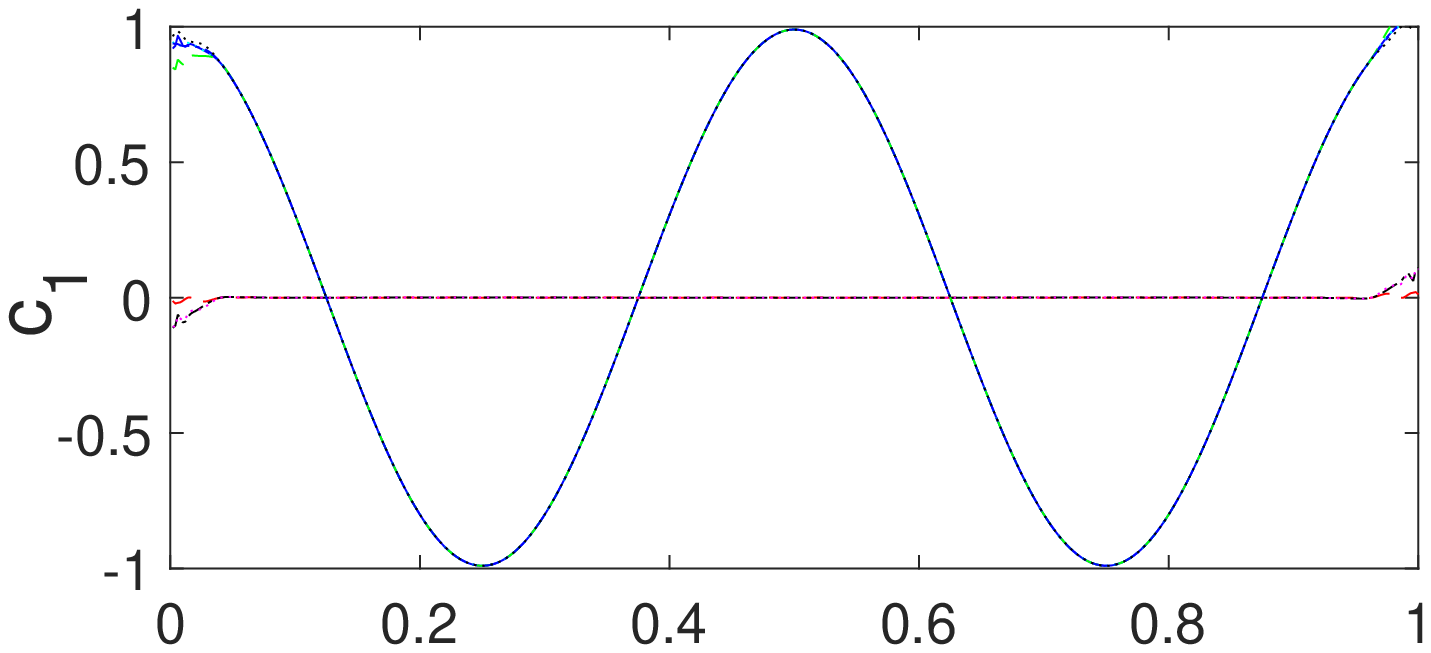}
\includegraphics[width=0.25\textwidth]{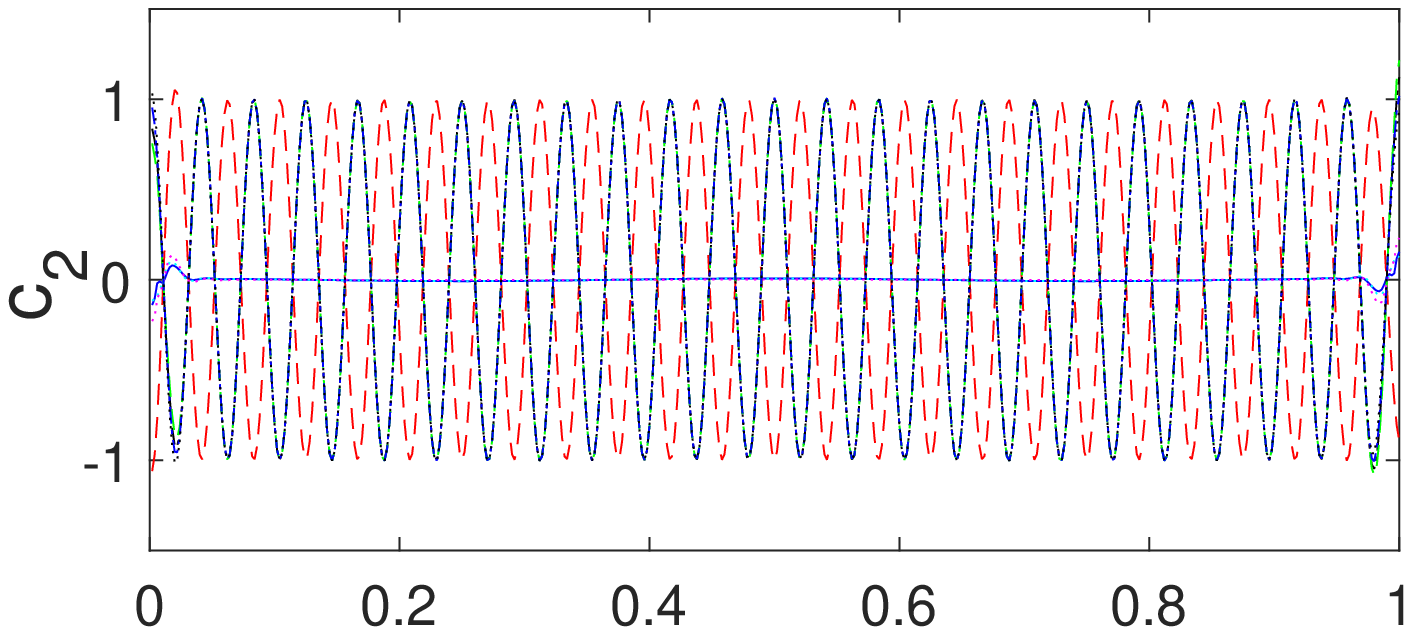}\hspace{-4mm}
\includegraphics[width=0.25\textwidth]{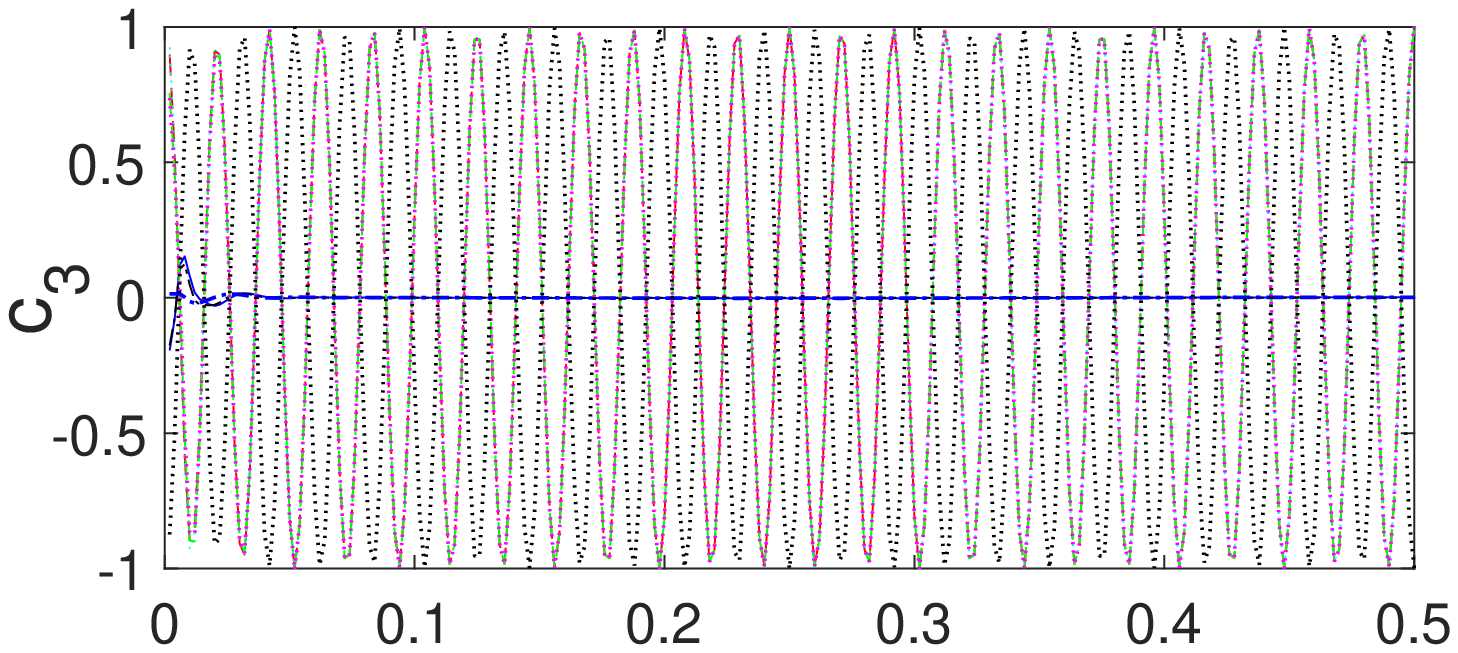}
\centering
\includegraphics[width=0.255\textwidth]{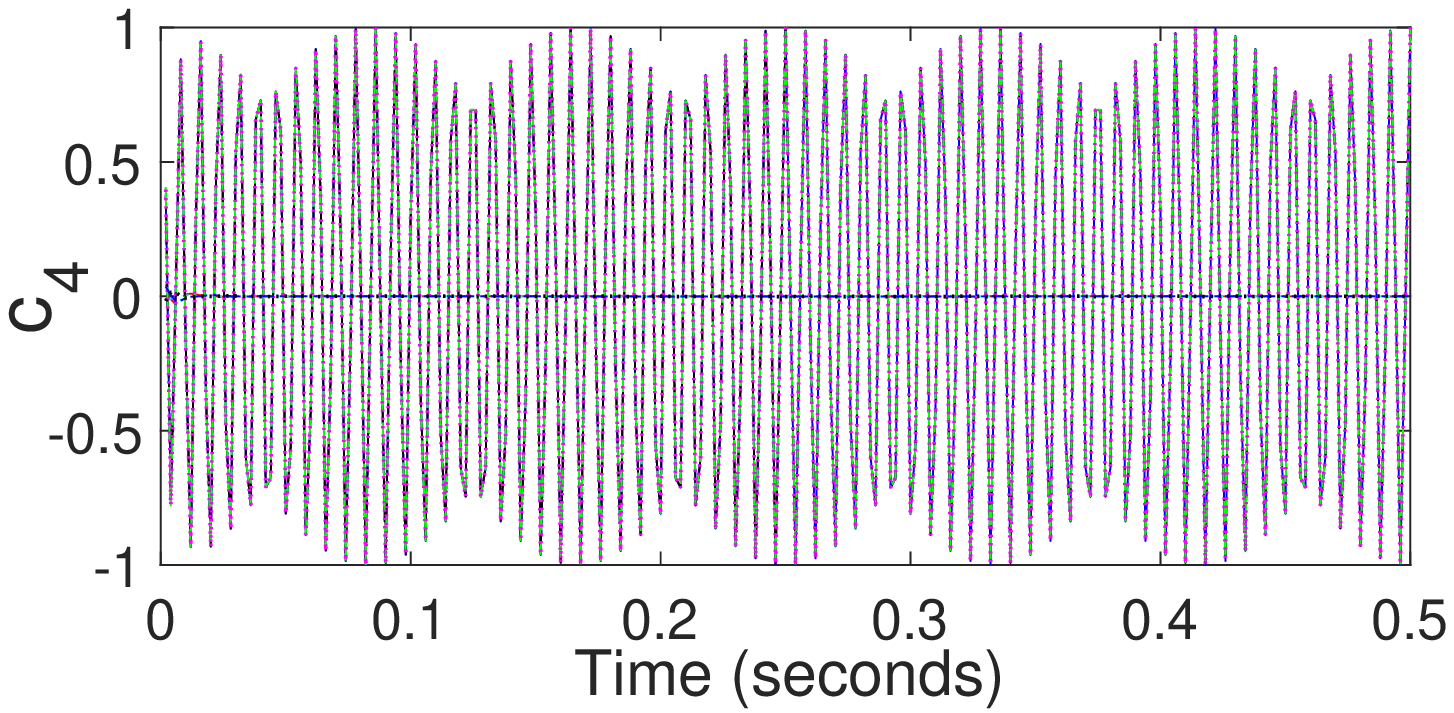}
\caption{Time-series plots of input synthetic signal and decomposed components obtained by applying the TVGMD method.}
\label{fig:syn1}
\end{figure}

\begin{figure}[h]
\begin{subfigure}{.5\textwidth}
  \centering
  \includegraphics[width=.85\linewidth]{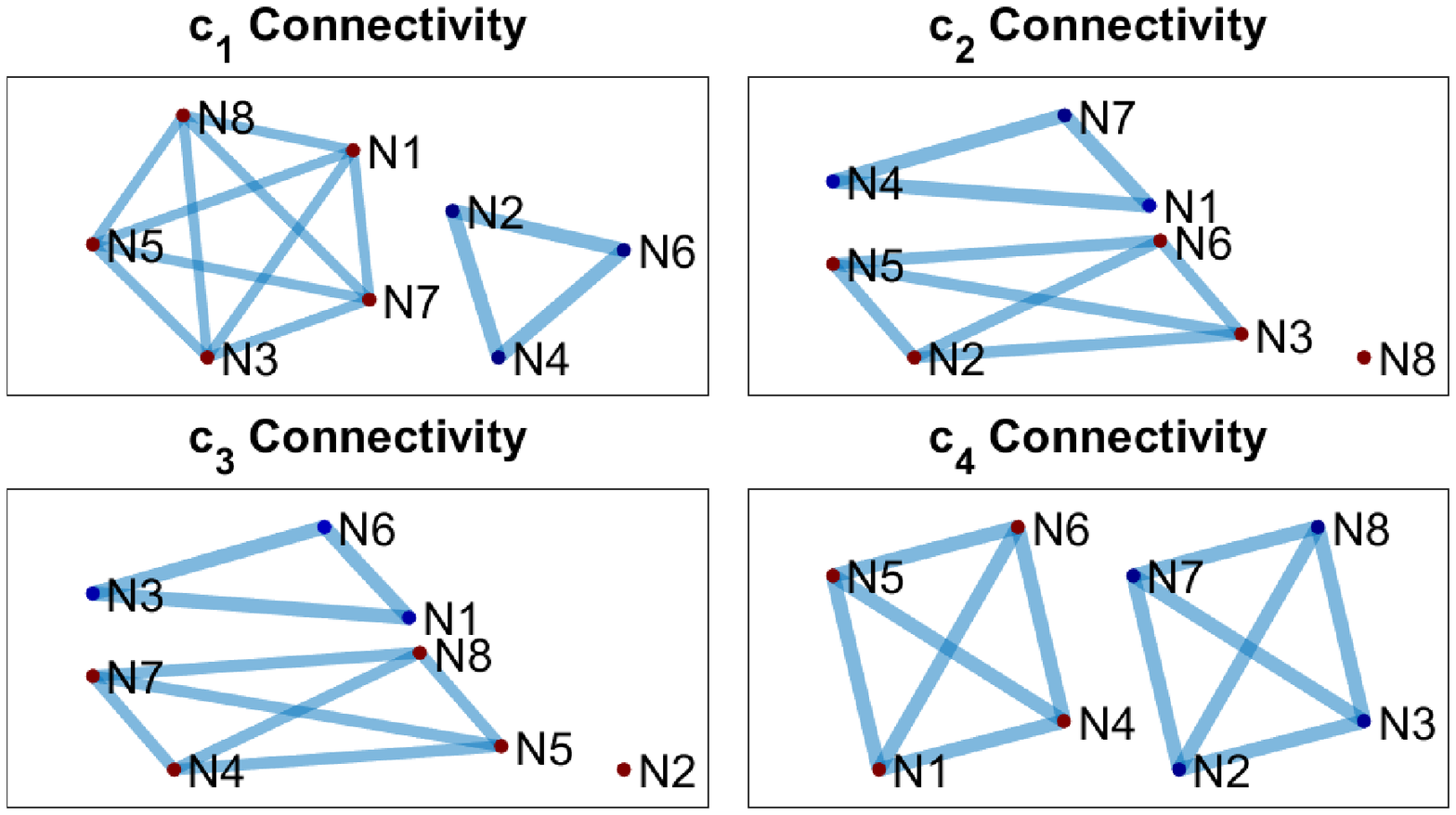}  
  \caption{}
  \label{fig:sub-first}
\end{subfigure}
\begin{subfigure}{.5\textwidth}
  \centering
  \includegraphics[width=.85\linewidth]{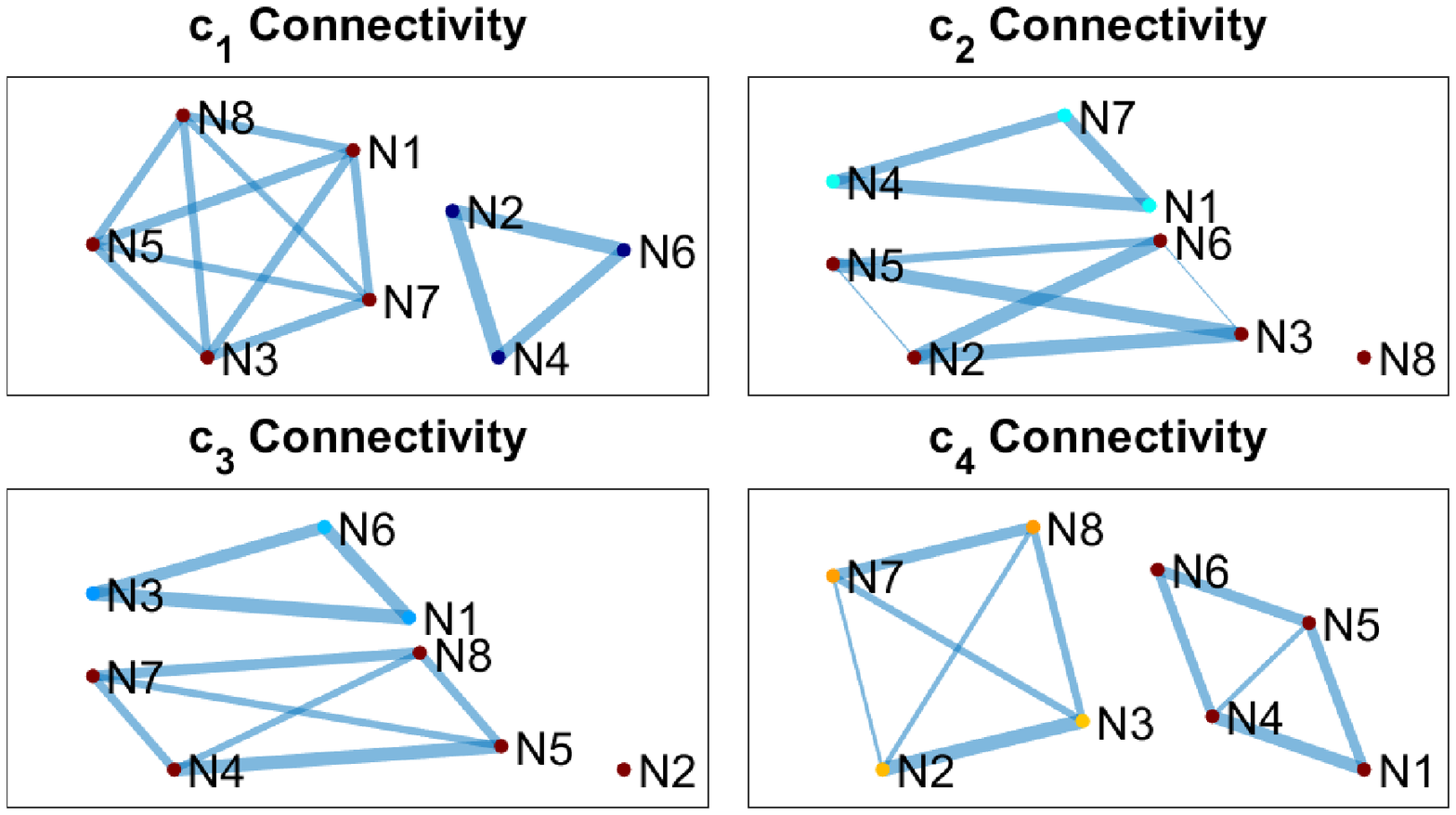}  
  \caption{}
  \label{fig:sub-first}
\end{subfigure}
\caption{Multi-scale connectivity matrices obtained by applying the TVGMD method on a) clean input signal \eqref{eq:syn}; b) noisy input signal \eqref{eq:syn} with the value of $\eta_i$ that correspond to the $SNR=6 \mbox{ }dB$ for all $x_i$.}
\label{fig:syn2}
\end{figure}

\subsection{Analysis of Time-varying Synthetic Graph Signal}
The first experiment involves a time-varying synthetic graph signal consisting of 8 nodes with each containing a combination of oscillations. The goal of this experiment is to demonstrate that the graph modes obtained from the TVGMD algorithm exhibit the following properties: i) the graph modes comprise band-limited time-varying signal oscillations; ii) for each graph mode, the corresponding adjacency matrix contains information about the connectivity structure of the mode. Further, the noise robustness of the TVGMD algorithm will also be assessed on the input data. Owing to the synthetic nature of the data, the ground truth for each graph mode is readily available that will facilitate the assessment of the TVGMD method.  In this experiment, the parameter values of the TVGMD method were: $\alpha=200$, $\beta=0.1$, $\gamma=1$, $\tau=0$ and $K=4$. %

The input time-varying graph signal, denoted by $\mathbf{x}=[x_1, x_2, \ldots, x_8]$, is defined as follows. 
\begin{align}
\centering
    x_1&=\cos(2\pi\times2t)+\cos(2\pi\times 128t)+\eta_1 \notag\\
    x_2&=-\cos(2\pi\times24t)+\cos(2\pi\times 48t)+\eta_2 \notag\\
    x_3&=\cos(2\pi\times 2t)+\cos(2\pi\times 48t)+\eta_3 \notag\\
    x_4&=\cos(2\pi\times 24t)+\cos(2\pi\times 128t)+\eta_4 \notag\\
    x_5&=\cos(2\pi\times 2t)+\cos(2\pi\times 24t)+\cos(2\pi\times 48t)\nonumber\\
    & \quad +\cos(2\pi\times 128t)+\eta_5 \notag \\
    x_6&=\cos(2\pi\times 48t)+\cos(2\pi\times 128t)+\eta_6 \notag\\
    x_7&=\cos(2\pi \times2t)+\cos(2\pi\times 24t)+\eta_7 \notag\\
    x_8&=\cos(2\pi\times 2t)+\cos(2\pi\times 24t)-\cos(2\pi\times 48t)+\eta_8
    \label{eq:syn}
\end{align}
Note that each of the graph nodes consists of a combination of up to four oscillatory signals: 2 Hz, 24 Hz, 48 Hz and 128 Hz sinusoid. $\eta_i$ denotes the additive Gaussian noise added to the signal associated with the $i$-th graph node. The oscillations in the graph signal \eqref{eq:syn} also defines its connectivity structure at each scale. For instance, as the 2 Hz oscillation is commonly present within the signals at graph nodes $[x_1,x_3,x_5,x_7,x_8]$, we expect those nodes to be connected at the corresponding (low-frequency) scale. At the same scale, the remaining nodes $[x_2,x_4,x_6]$ would also be connected as their corresponding time-series are all close to zero.       

The time series plots of the original noiseless graph signal (for $\eta_i=0$) as well as the decomposed graph modes obtained from the TVGMD method are shown in Fig. \ref{fig:syn1}. Here, the modes $c_1$, $c_2$, $c_3$ and $c_4$ are associated with the oscillations of frequencies 2 Hz, 48 Hz, 24 Hz and 128 Hz respectively. In each case, the time series plots for all graph nodes ($N1$-$N8$) are depicted in the Fig. \ref{fig:syn1} in different colours. 

Fig. \ref{fig:syn2}(a) shows the corresponding connectivity structure of each graph mode, inferred from the corresponding adjacency matrix. Each node is depicted by a red or a blue marker in the connectivity graphs, with the red (blue) node indicating that the corresponding graph mode has high (low) signal energy.

Note that there is a complete alignment of information across different scales in the extracted graph modes as shown in Fig. \ref{fig:syn1}; for instance, only 2 Hz oscillation is present in $c_1$; 48 Hz in $c_2$; 24 Hz in $c_3$ and 128 Hz in $c_4$. This alignment of information is a crucial requirement in many signal processing applications involving multivariate signals e.g., data fusion \cite{Rehman15} and denoising \cite{Rehman2019}. 

\begin{figure*}[t]
\includegraphics[width=0.355\textwidth]{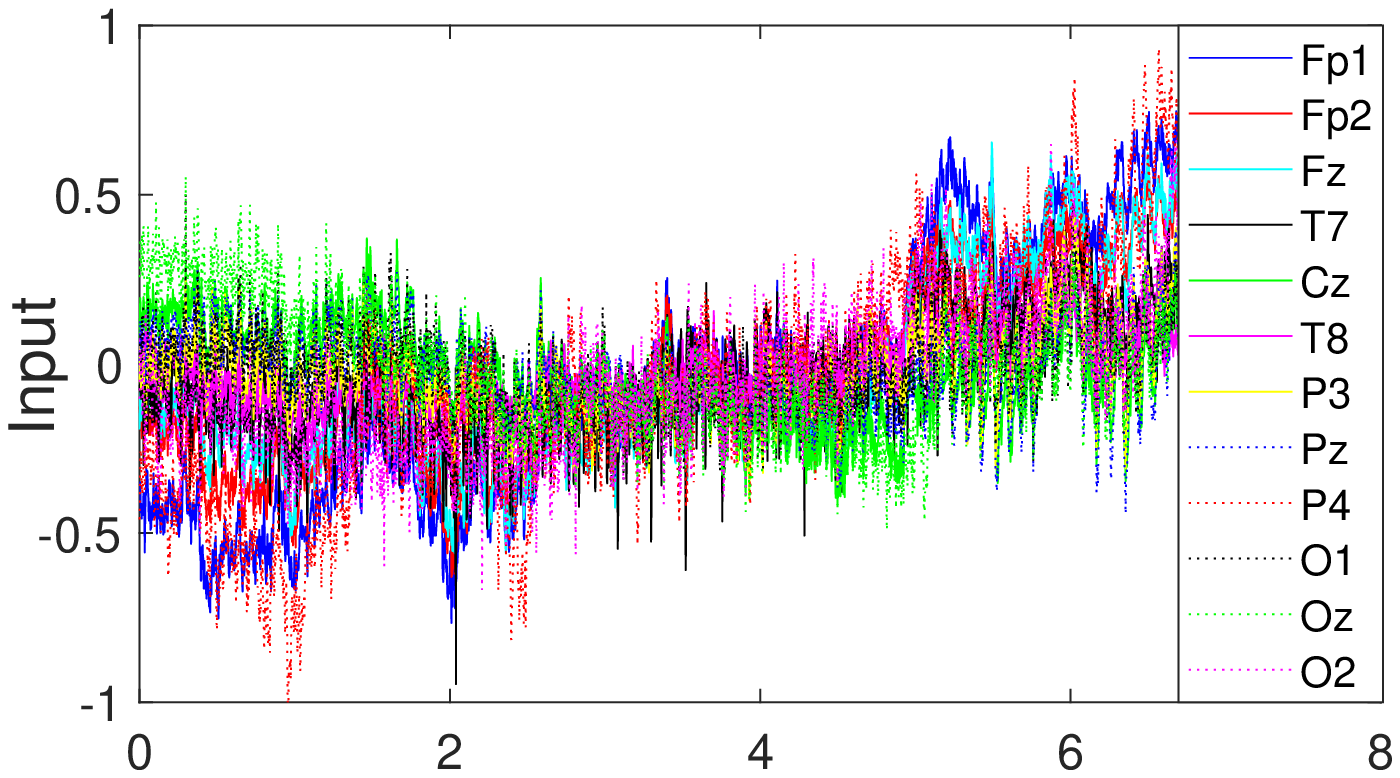} \hspace{-6mm}
\includegraphics[width=0.355\textwidth]{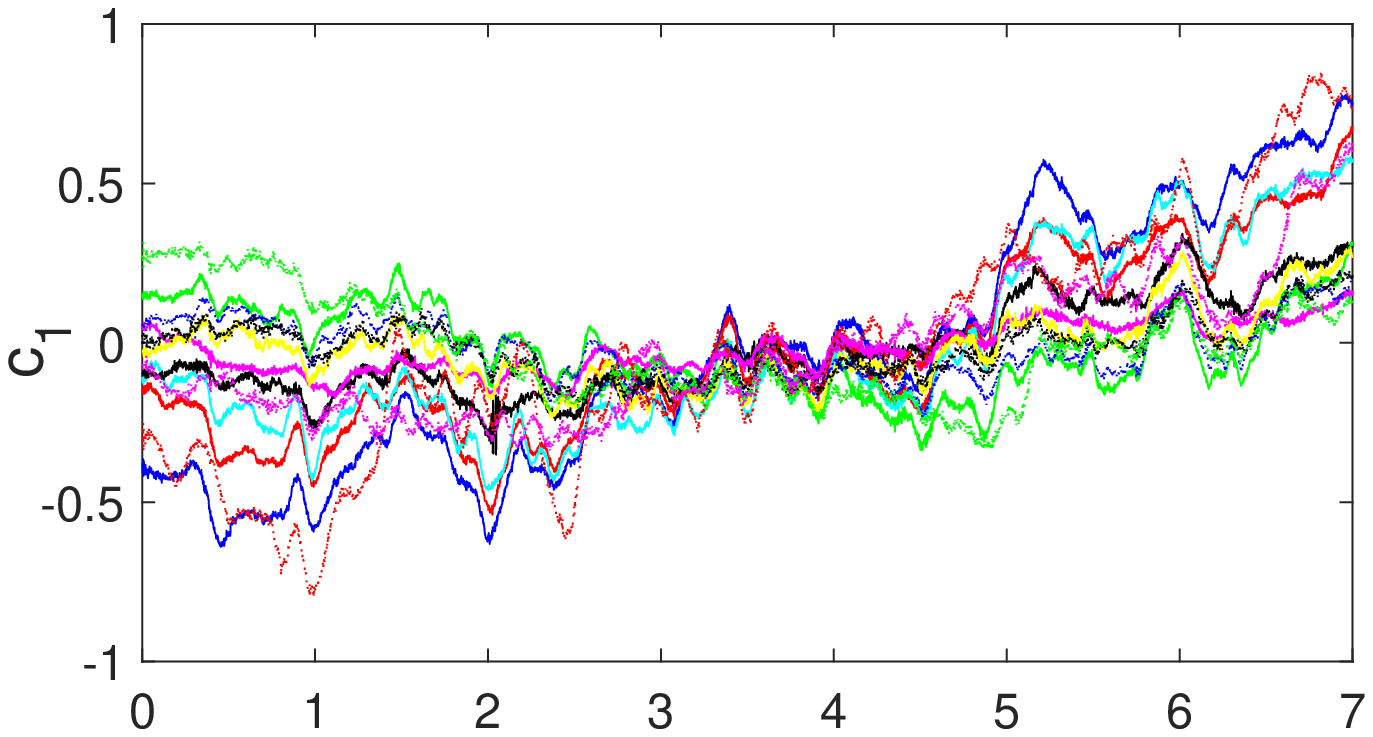} \hspace{-6mm}
\includegraphics[width=0.355\textwidth]{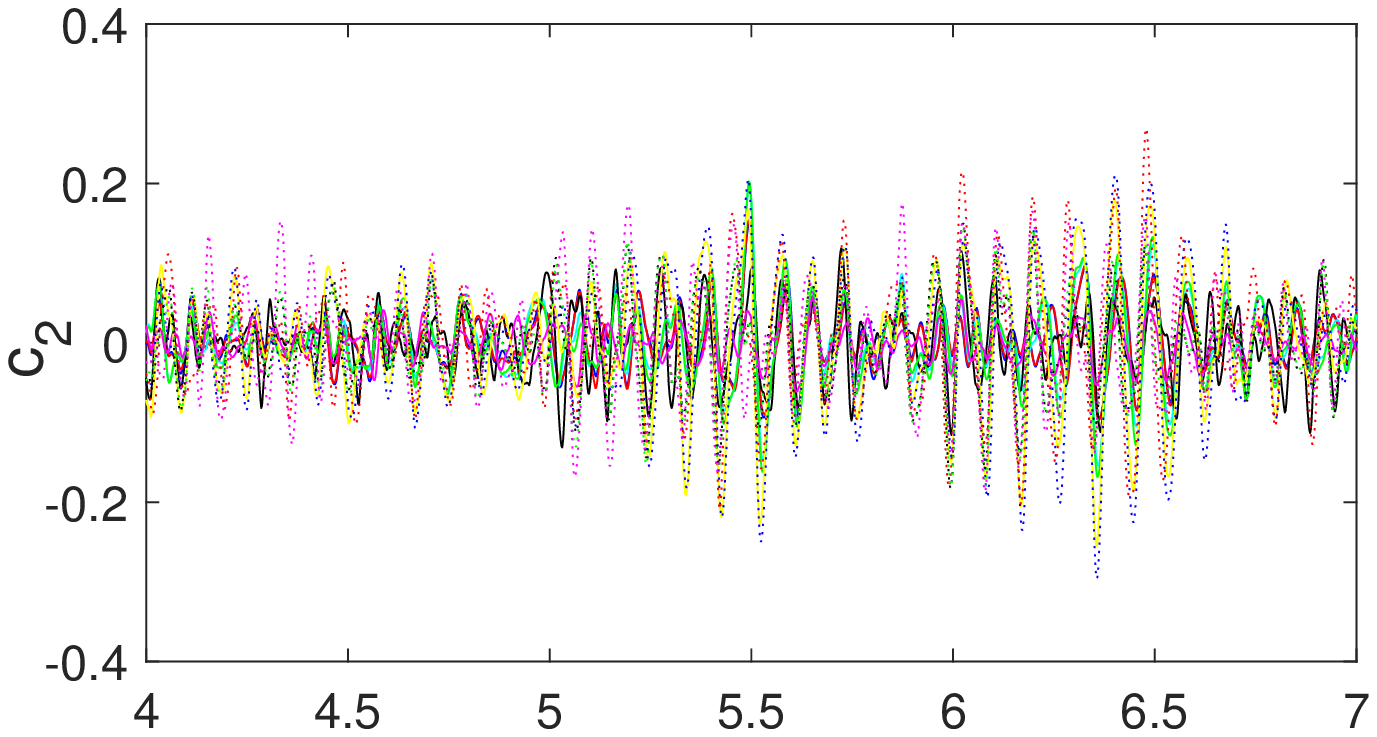}\\
\includegraphics[width=0.355\textwidth]{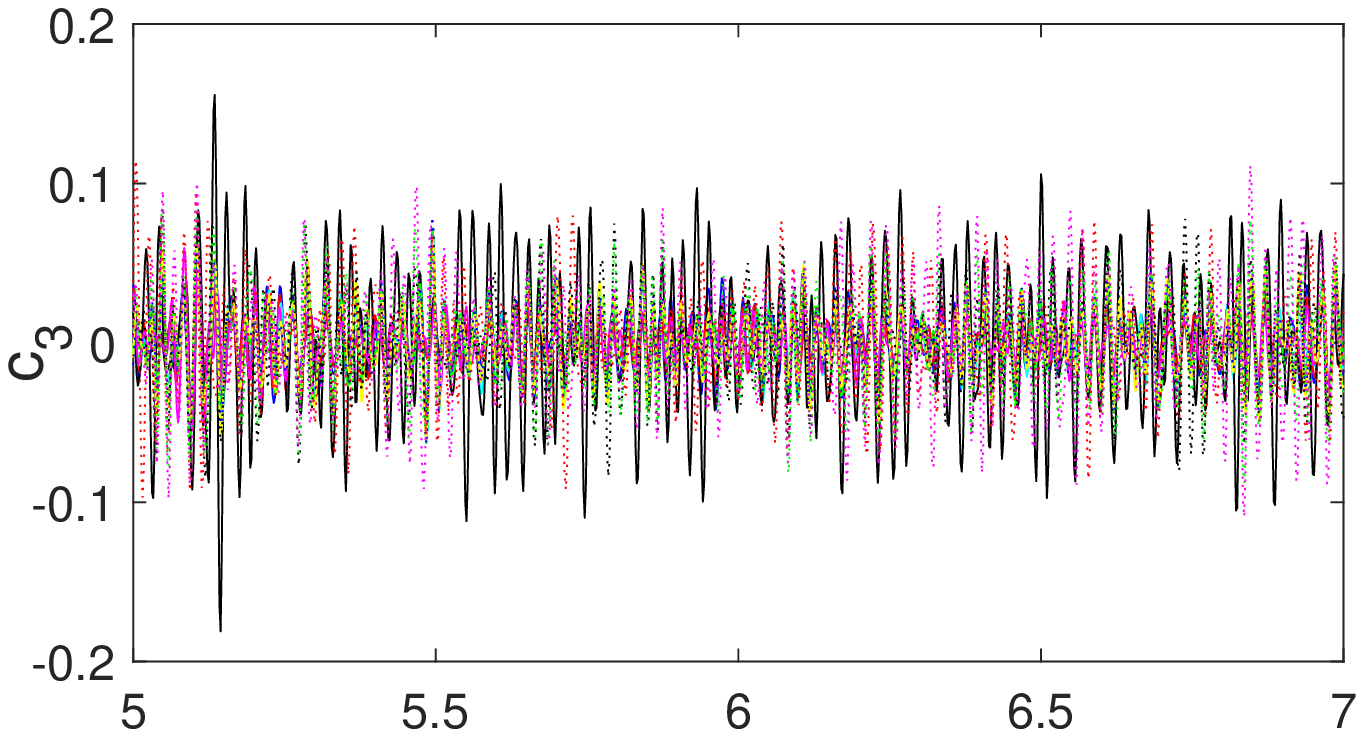} \hspace{-6mm}
\includegraphics[width=0.355\textwidth]{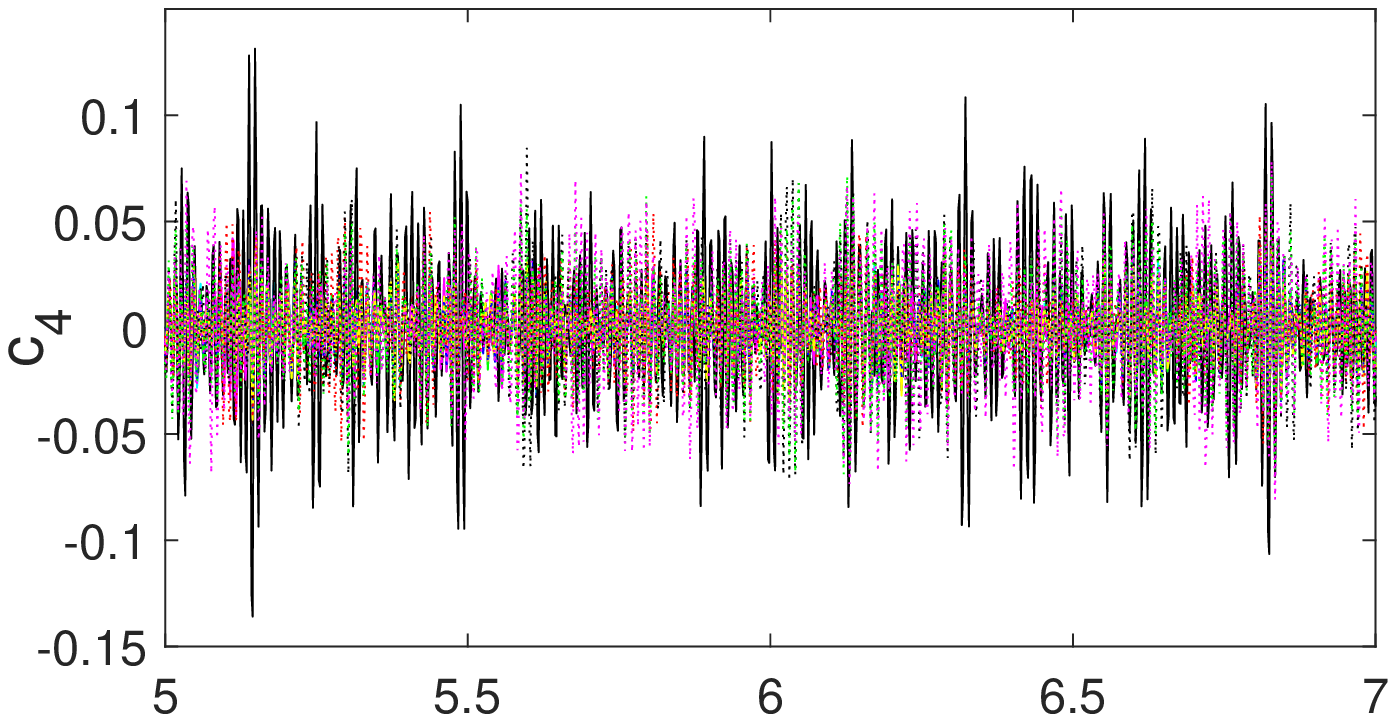} \hspace{-6mm}
\includegraphics[width=0.355\textwidth]{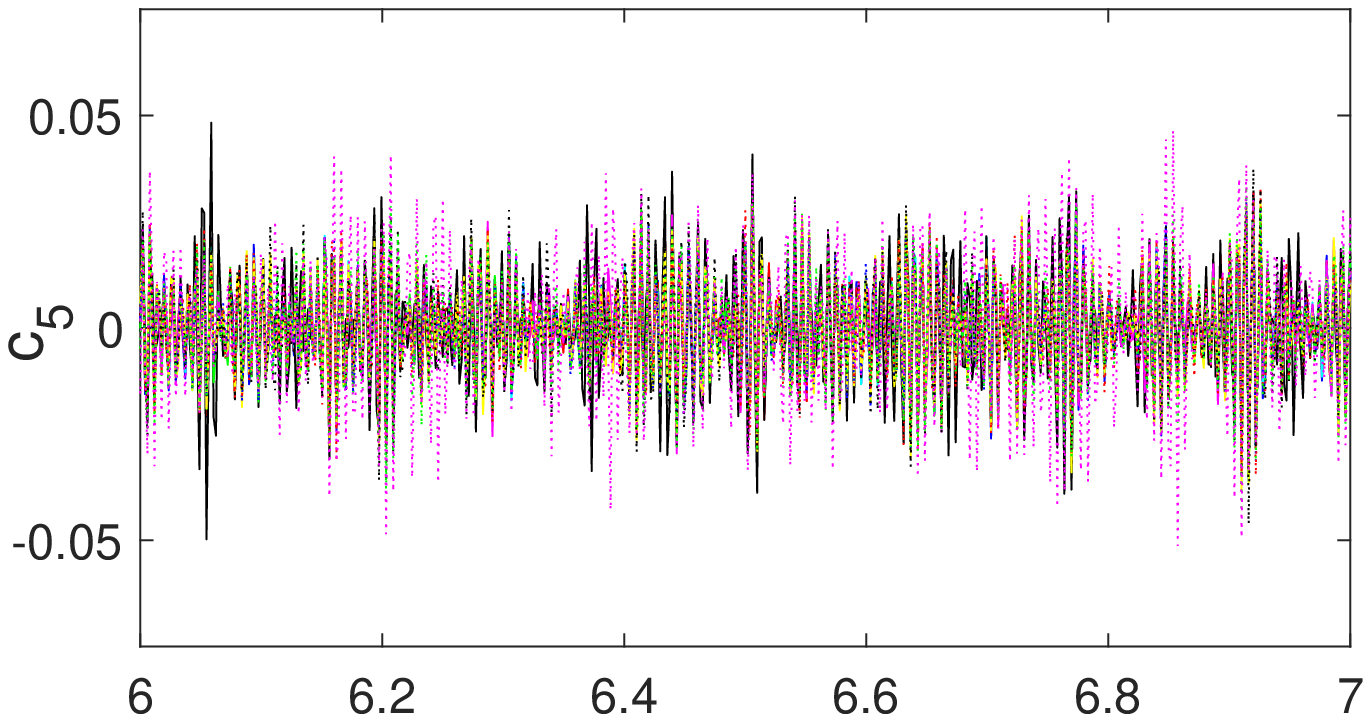}\vspace{2mm}
\caption{ Time series plots of the 12-channel input EEG signal and the decomposed components obtained by using the TV-GMD algorithm. The x-axis denotes the number of samples; the sampling frequency was 512 Hz. EEG amplitudes were scaled in the range of $[-1,1]$ for better illustration.}
\label{fig:eeg_ts}
\end{figure*}

The analysis of the adjacency matrices of the graph modes in Fig. \ref{fig:syn2}(a) also reveals the underlying connectivity patterns in the data. For example, the connectivity pattern of the first graph mode $c_1$ suggests that the high-energy nodes $N1$, $N3$, $N5$, $N7$ and $N8$ are connected, whereas $N2$, $N4$ and $N6$ are close to zero (hence the blue colored nodes) and are thus connected to each other. This observation is consistent with how the graph signal has been defined in \eqref{eq:syn}. The same could be concluded for the other graph modes: in $c_3$ for instance,  the nodes $N4$, $N5$, $N7$ and $N8$ carry the 24 Hz oscillation signals and are correctly shown as connected, whereas $N1$, $N3$ and $N6$ are all close to zero and therefore also connected. An interesting case here is that of node $N2$ that is not connected to the remaining nodes. The reason is that the term associated with the 24 Hz sinusoid in $N2$ (or $x_2$) \eqref{eq:syn} has negative sign meaning that it is out-of-phase with the 24 Hz oscillations at the other nodes. 

Finally, the connectivity graphs obtained for the noisy graph signal \eqref{eq:syn} are shown in Fig. \ref{fig:syn2}(b). The noise variance  at each graph node was set to a level that resulted in the $SNR=6\mbox{ }dB$ for each of the input signal $x_i$. Comparing these connectivity graphs with those obtained for the clean signal in Fig. \ref{fig:syn2}(a), it is clear that the addition of noise affects the output of the TVGMD method and hence alters the connectivity patterns. That said, the primary data patterns at multiple scales remain intact. Specifically, for $c_2$, we observe a partially-connected graph between $N2$, $N3$, $N5$ and $N6$ instead of a fully connected one. Similar observations can be made regarding the connectivity patterns of $c_4$. While this result gives some insight into the performance of the TVGMD in the presence of noise, a detailed study on the topic could be an avenue for future work.

\begin{figure}[t]
\includegraphics[width=0.55\textwidth]{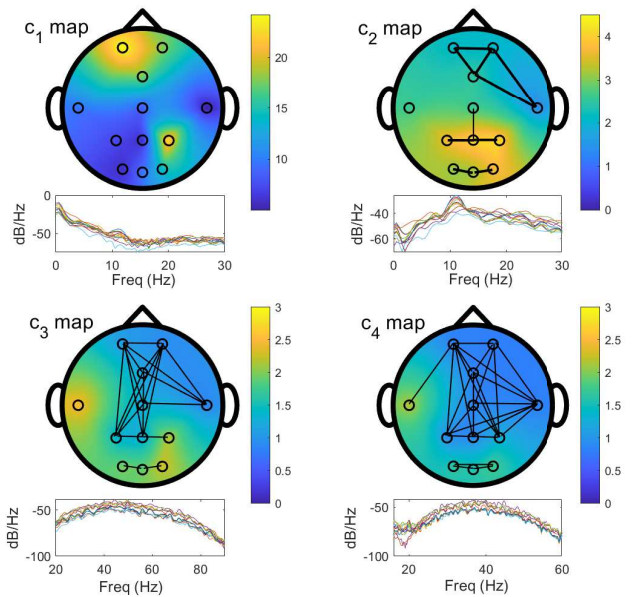}\\
\caption{ Connectivity graphs of the EEG data at multiple scales obtained from TVGMD. No connectivity exists among the nodes at $c_1$, whereas strong connectivity exists among the nodes representing the occipital and the parietal regions within $c_2$. This makes sense as $c_2$ carries alpha rhythms which are typically dominant in the occipital and parietal brain regions.}
\label{fig:eeg_graph}
\end{figure}

\begin{figure}[t]
\centering
\includegraphics[width=0.51\textwidth]{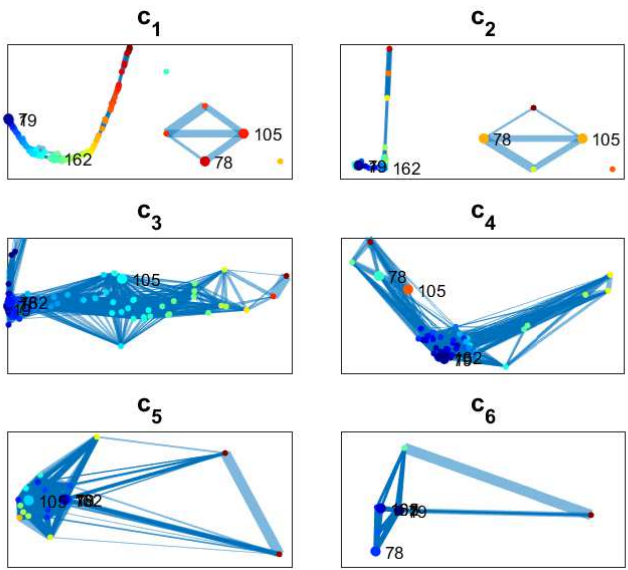}
\caption{ Multi-mode connectivity graphs of the electricity consumption time-varying graph signal, obtained via TVGMD.}
\label{fig:electricity1}
\end{figure}

\subsection{Analysis of Alpha Waves in Electroencephalogram (EEG)}
This experiment will demonstrate the potential of the TVGMD method in the analysis of biomedical electroencephalogram (EEG) signals. EEG is a recording of brain's electrical activity, obtained from multiple sensors (electrodes) attached to the scalp. Typically, each electrode records a time-series of electrical signal produced by brain. This way, multichannel EEG signals can be represented by time-varying graph signals with each electrode representing a graph node. 

EEG signals are known to exhibit oscillatory behaviour in specific frequency bands e.g., theta (4-8Hz), alpha (8-12 Hz), beta (13-30 Hz) and gamma (30-150 Hz) bands, each linked with different underlying neural processes. Therefore, a crucial task in numerous applications involving EEG signals is to decompose neural oscillations into specific bands. Further, it is of interest to reveal the functional connectivity structures in brain at multiple frequency bands. Here, we demonstrate the potential of the TVGMD method to concurrently provide both pieces of information i.e., time-varying modes and their connectivity structures.        

Here, we apply the TVGMD method to the publicly available EEG data that was obtained in a resting-state eyes-open/closed experimental protocol. The data was recorded in the GIPSA-lab, Grenoble, France, in 2017. A research-grade amplifier (g.USBamp, g.tec, Schiedlberg, Austria) along with an EC20 cap comprising 16 electrodes, placed according to the 10-20 international system, was used to collect the data. Further details about the data acquisition process and the related experiment can be found in  \cite{gregoire18}. The TVGMD parameter values used in this experiment were: $\alpha=1000$, $\beta=0.5$, $\gamma=1$, $\tau=0.1$ and $K=6$.

The results of applying the TVGMD algorithm on the 12-channel EEG signal is shown. Particularly, the input and the time-series associated with the obtained graph modes are shown in the Fig. \ref{fig:eeg_ts} while the corresponding functional connectivity structures are shown in the Fig. \ref{fig:eeg_graph}. Note that the time plots of the first mode ($c_1$) mainly contains the low-frequency trend of the signal. A look at the corresponding functional connectivity map of $c_1$ in the lower figure reveals no prominent connections among different brain regions as expected. The power spectrum of the graph modes corresponding to each node are shown in the inset plot with each connectivity graph.

The time series of the second graph mode, $c_2$, comprises oscillations related to the alpha band which are of particular interest here since the EEG data was obtained in the eyes-closed state, where alpha rhythms are typically prominent in the occipital and the parietal regions. This can be verified in the corresponding connectivity map of $c_2$ that not only shows consistently higher signal power in the occipital and the parietal regions but also a strong functional connectivity within that region. Note that in the corresponding Fourier spectra of the graph mode (shown in the inset of the connectivity graph), there is a peak at the 8-12 Hz frequency range which implies that we are indeed operating in the alpha frequency band.   

For the remaining couple of graph modes, $c_3$ and $c_4$, the time plots and the connectivity patterns are also shown. For these modes, we observe mostly low-energy signals as well as random connectivity patterns in the different brain regions.  

\begin{figure}[t]
\includegraphics[width=0.255\textwidth]{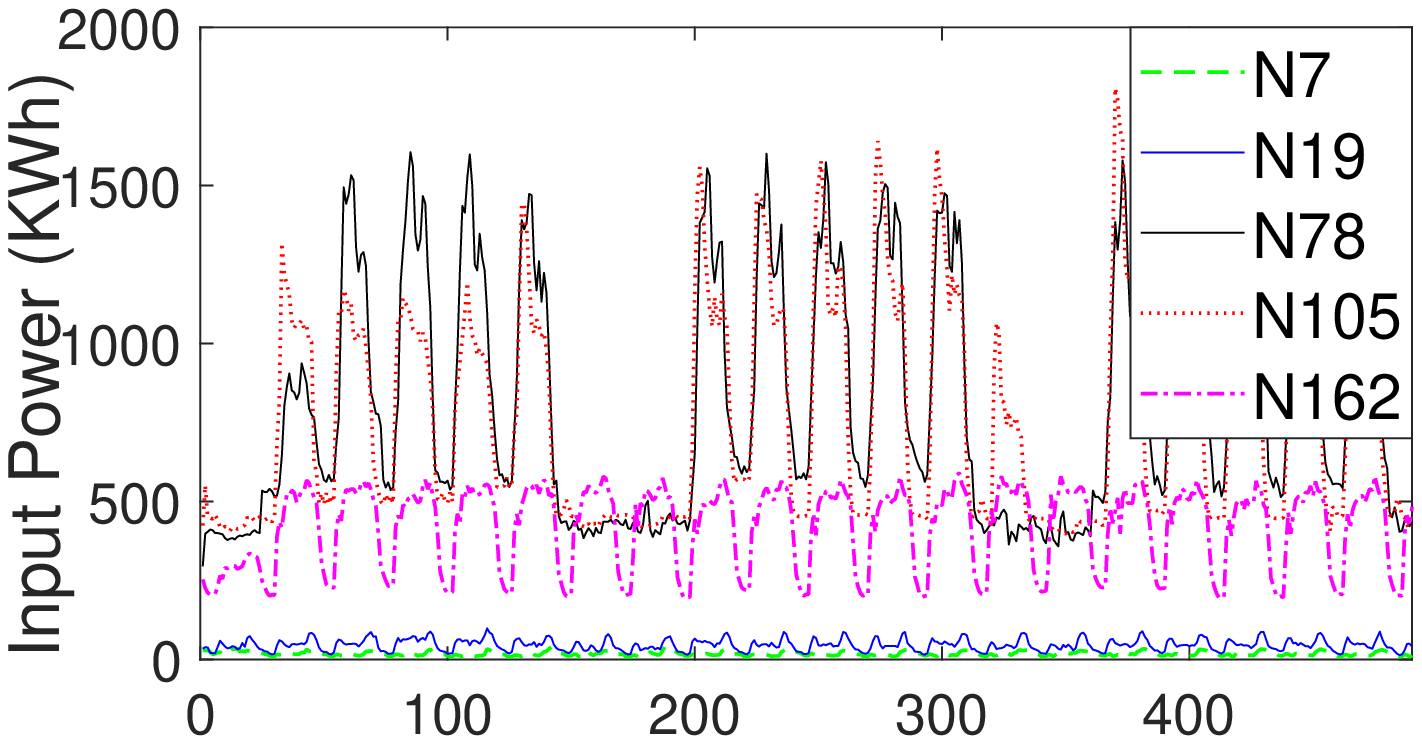} \hspace{-6mm}
\includegraphics[width=0.255\textwidth]{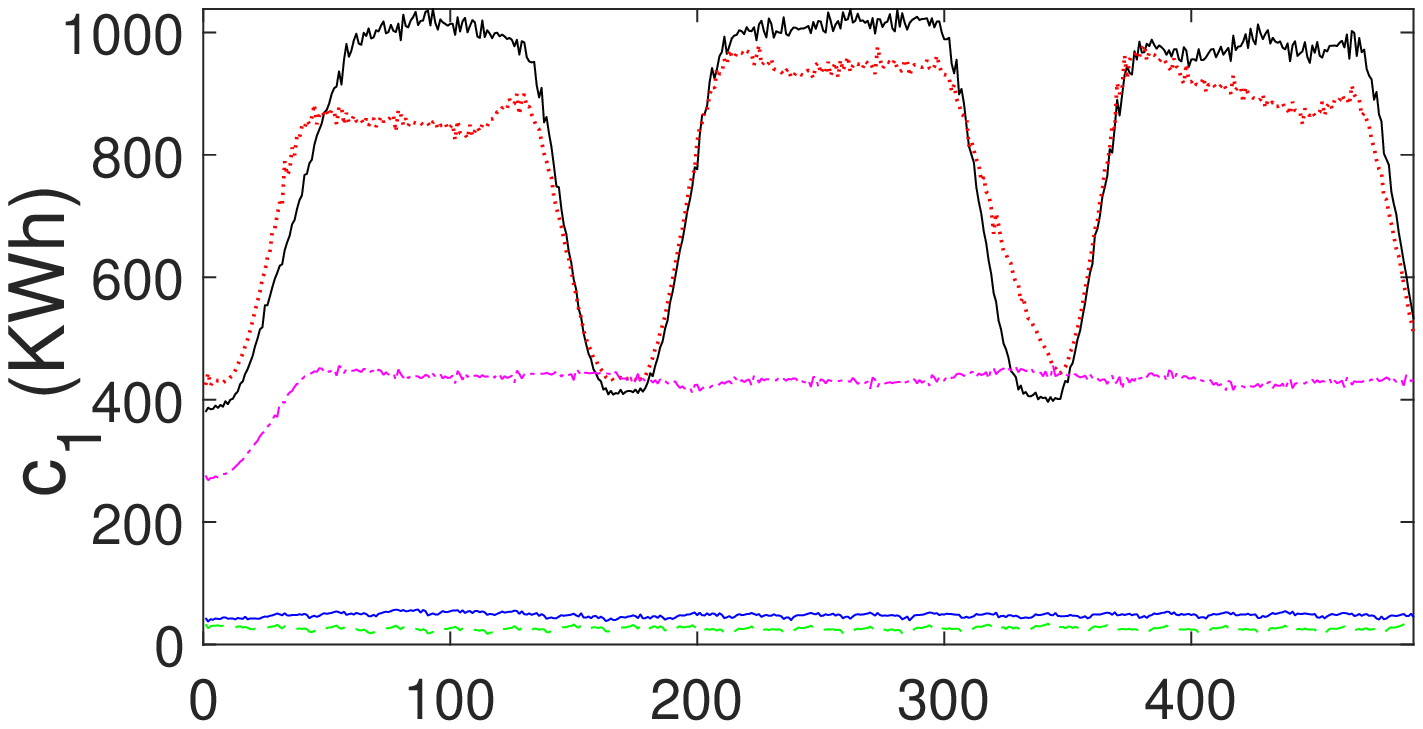}\hspace{-6mm} \\
\includegraphics[width=0.255\textwidth]{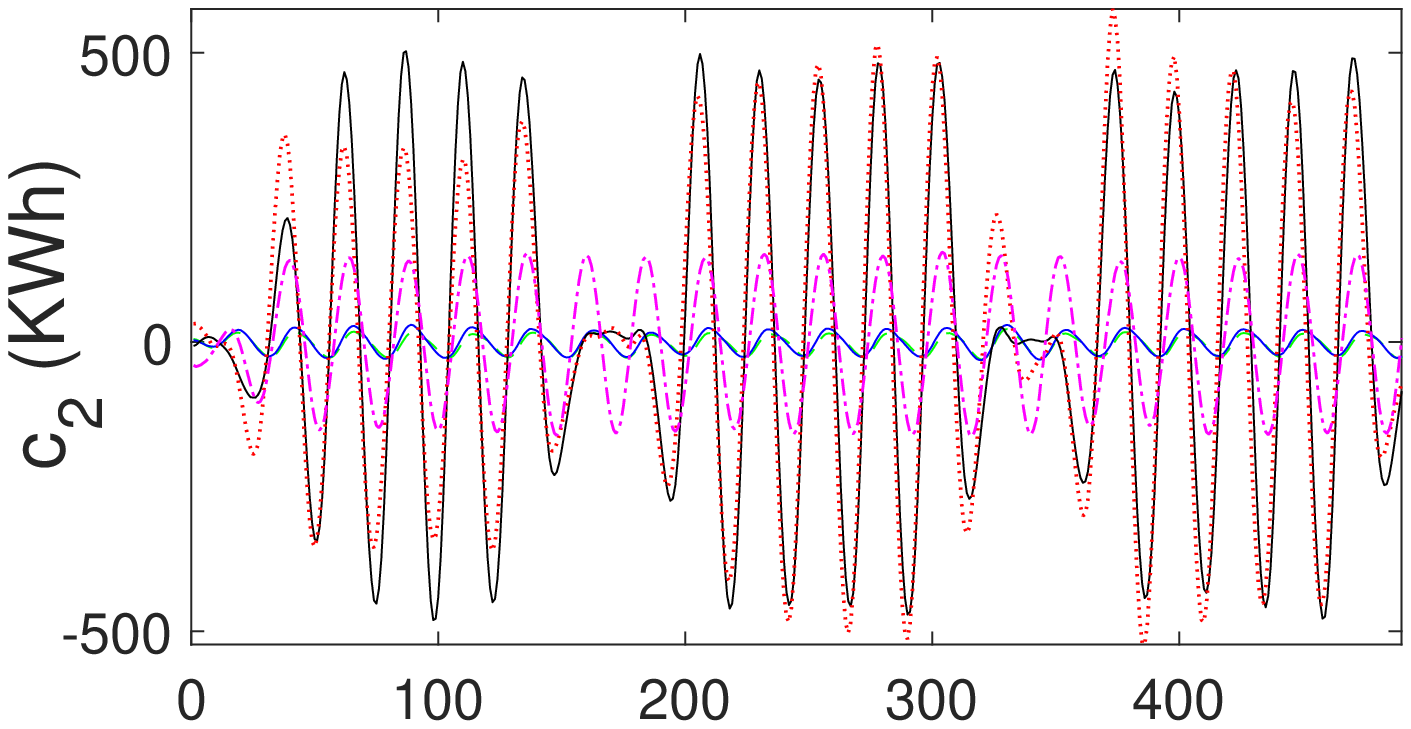}\hspace{-5mm}
\includegraphics[width=0.255\textwidth]{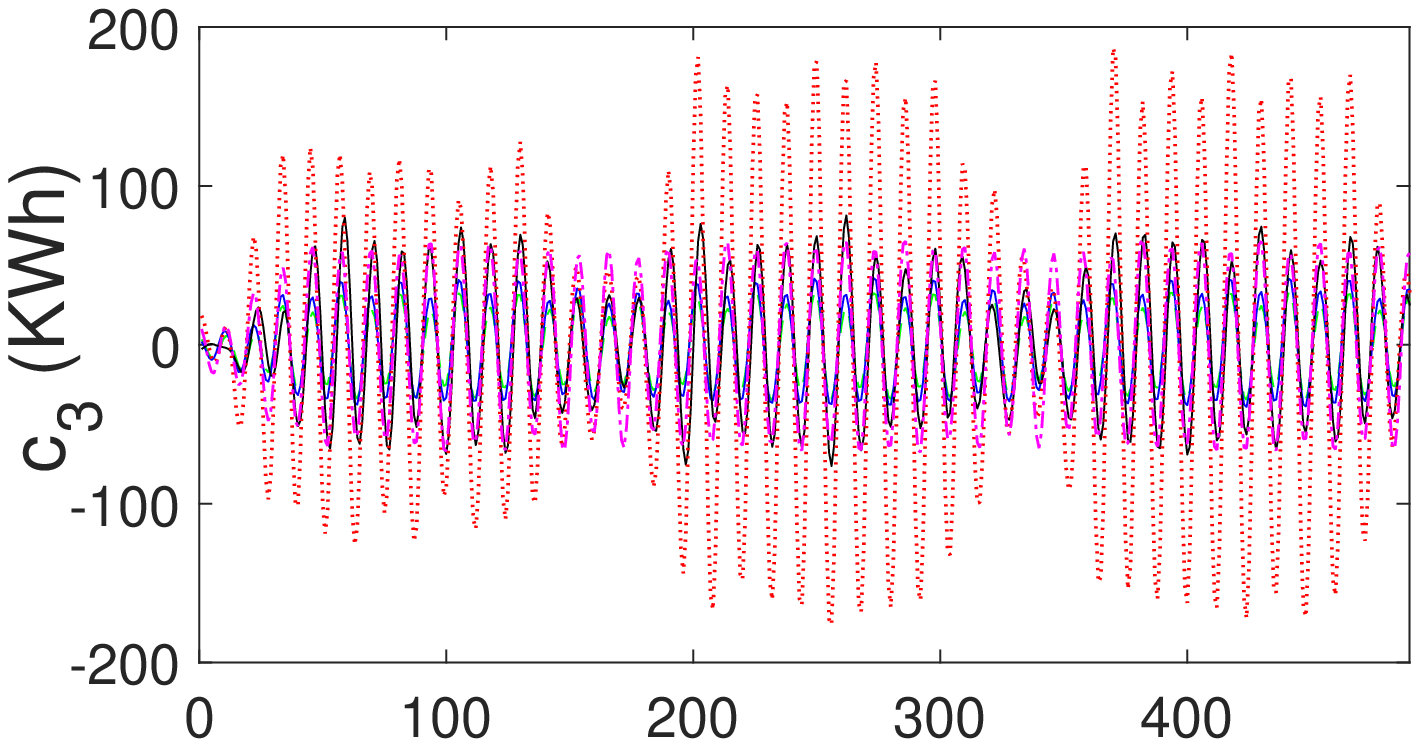}\hspace{-6mm} \\
\includegraphics[width=0.255\textwidth]{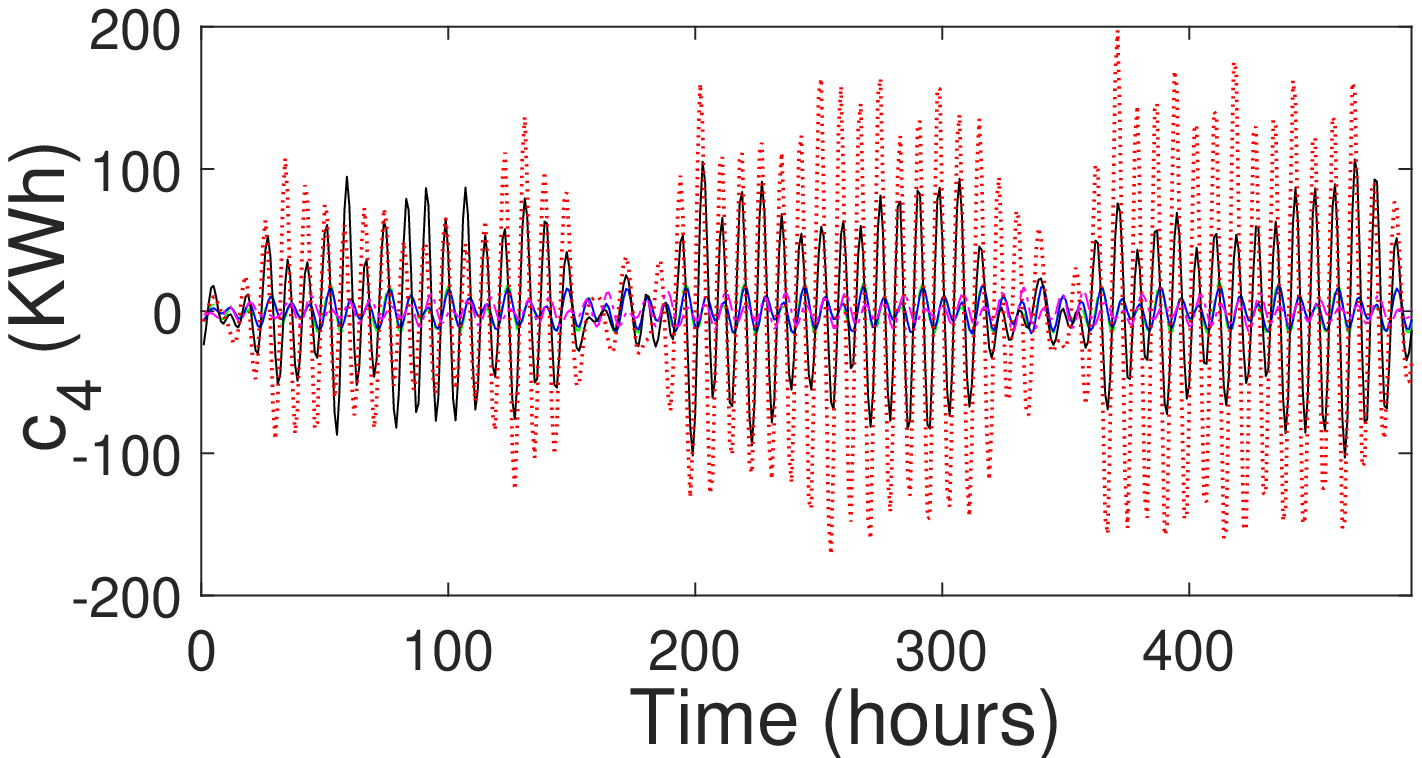} \hspace{-6mm}
\includegraphics[width=0.255\textwidth]{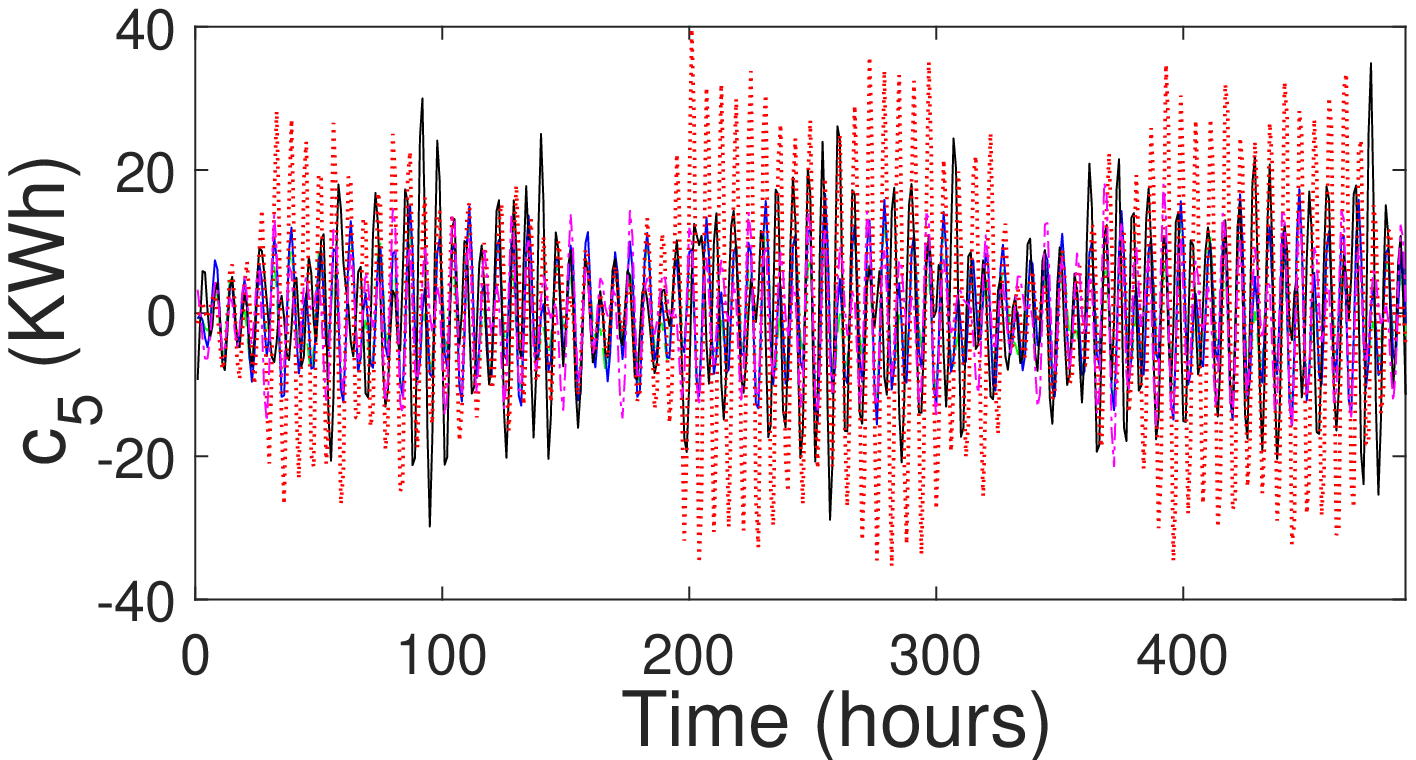}\hspace{-6mm}
\caption{Time-series plots of the multiple graph modes of the electricity consumption data at five selected node.}
\label{fig:electricity2}
\end{figure}
\subsection{Electric Consumption Data Analysis in Power Networks}
The next time-varying graph data set used in our experiment is the electricity consumption in kWh, recorded every 15 minutes from 2012 to 2014 in Portugal for 206 clients \cite{Dua19}. The data was transformed to show hourly consumption of electricity and 480 hours (20 days) of data was analyzed using the TVGMD method; the results are shown in Fig. \ref{fig:electricity1} and Fig. \ref{fig:electricity2}. The following TVGMD parameter values were used in this experiment: $\alpha=2000$, $\beta=0.25$, $\gamma=1$, $\tau=0$ and $K=10$.

Fig. \ref{fig:electricity1} shows the connectivity structures of the first 6 graph modes obtained from the TVGMD method. Note that 5 selected clients (nodes) have been specifically highlighted in the connectivity graphs: N7, N19, N78, N105 and N162. The electricity consumption time-series plots of the graph modes for these 5 clients (nodes) along with the overall consumption are shown in Fig. \ref{fig:electricity2}. These nodes have been carefully selected out of the total 206 nodes to show a variety of consumption patterns, ranging from a high consumption (at Nodes N78 and N105), mid-range consumption (at the Node N162) to a very low consumption (at Nodes N7 and N19).   

By analyzing the graph modes in time domain, it was noticed that the electricity consumption followed oscillatory patterns at different time scales that were extracted accurately through the obtained modes. Specifically, $c_5-c_1$ provided information about the 6-hourly, 8-hourly, 12-hourly, daily and 1-weekly electricity consumption patterns respectively. The adjacency graphs gave further information regarding the connectivity of different nodes (clients) at different time scales in terms of their electricity usage: for example, the $c_1$ and $c_2$ time plots show that the nodes 78 and 105 follow similar consumption patterns on weekly and daily (24 hourly) basis which can be verified through their strong links in the $c_1$ and $c_2$ connectivity graphs. The nodes N7 and N19 also exhibited strong connections (in terms of electricity usage) across a range of different time scales that could be verified from the respective connectivity graphs (though those connections remain mostly hidden in our illustration due to their closeness and a very high density of the other nodes).  

\subsection{Detection of Plant-wide Oscillations in Flotation Circuits}
Flotation is used to separate useful mineral particles from gangue particles in a concentrator process \cite{Lindner18}. This is accomplished by imposing appropriate conditions that enable hydrophobic materials to attach to air bubbles and rise to the top of the flotation cell. Here, we consider data from a flotation cell that consists of two parallel banks with seven cells in series in each bank; the resulting signal can be modelled as a time-varying graph signal with 14 nodes. The process flow diagram of such a flotation cell is shown in Fig. 1 in \cite{Lindner18} along with the relevant description. The time-series data corresponding to the cell \textit{levels} and \textit{outflows} in the flotation circuit are of interest for analyzing the flotation plant process. Particularly, it is important to automatically detect oscillations in the plant data as well as diagnosing their root cause \cite{Chen21}. 
\begin{figure}[t]
\centering
\includegraphics[width=0.505\textwidth]{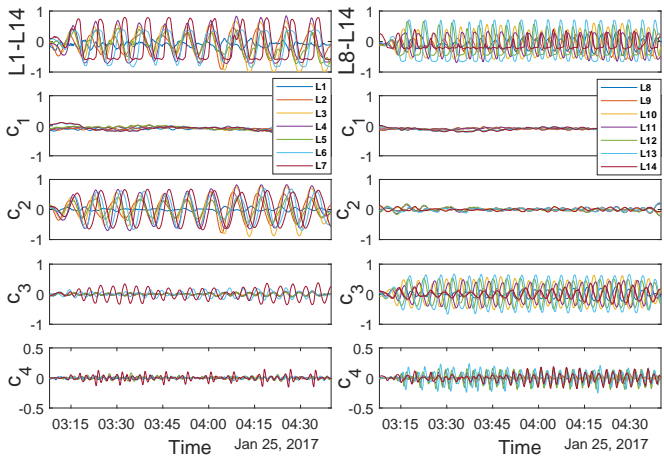}
\caption{ Time-series plots of the decomposed graph modes obtained by applying the TVGMD algorithm on the \textit{level} signals in a flotation circuit.}
\label{fig:float_time}
\end{figure}

\begin{figure}[t]
\includegraphics[width=0.51\textwidth]{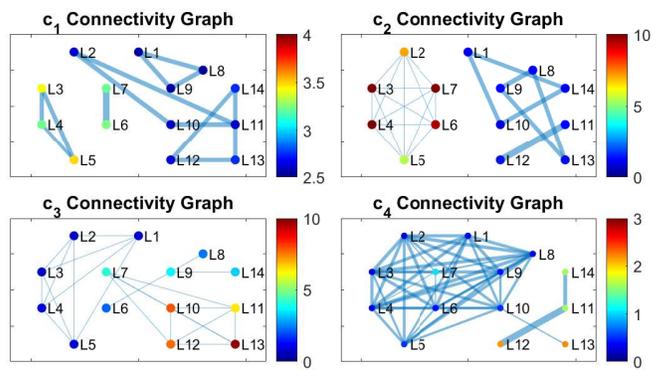}
\caption{Connectivity graphs corresponding to the graph modes from Fig. \ref{fig:float_time}. A natural grouping of the nodes (flotation cells) is evident from the $c_2$ and the $c_3$ connectivity graphs.}
\label{fig:float_graphs}
\end{figure}
The graph modes obtained from applying the TVGMD algorithm on input data are shown in Fig. \ref{fig:float_time} and Fig. \ref{fig:float_graphs}. The TVGMD parameter values used in this experiment were: $\alpha=1000$, $\beta=0.1$, $\gamma=1$, $\tau=0$ and $K=4$. The input data comprise the \textit{level} time-series, $L_1-L_{14}$, recorded from the 14 cells of the flotation circuit and is shown in the top row of Fig. \ref{fig:float_time}. The time plots of the decomposed graph modes clearly show the presence of two separate frequencies in the two flotation banks. Note that the lower frequency oscillation is prominent in the level time series of the first seven cells (in $c_2$), whereas the higher frequency oscillation is predominantly present in the remaining cells ($L_8-L_{14}$), in $c_3$. There is also a clear alignment of information present in the graph modes e.g., $c_2$ ($c_3$) contain only the specific low (high) frequency oscillation in all the cells or nodes of the graph signal. This \textit{mode-alignment} is a crucial requirement for the processing of multivariate signals in many application areas \cite{Rehman15}.

The corresponding connectivity graphs of the extracted graph modes are shown in Fig. \ref{fig:float_graphs}. The connectivity graphs corresponding to $c_2$ and $c_3$ are of particular interest here. The $c_2$ connectivity graph shows that the \textit{level} signals $L_2$-$L_7$ (from the first bank) are high-energy, as evident from the node colors, and are correlated with each other. The correlations are weak though (as evident from the thin connectivity links) owing to the inherent phase difference between the signals from different nodes. 
On the other hand, the connectivity graphs corresponding to $c_3$ show that the \textit{level} signals $L_9$-$L_{14}$ (from the second bank) have higher amplitudes and are correlated (albeit weakly) to each other. The connectivity graphs can therefore help infer relationships between measured variables at multiple scales. This could be used to devise grouping strategies for the diagnosis of root cause of oscillations in the flotation circuit, facilitating their fault diagnosis.

\section{Discussions}
\begin{figure}[t]
\includegraphics[width=0.25\textwidth]{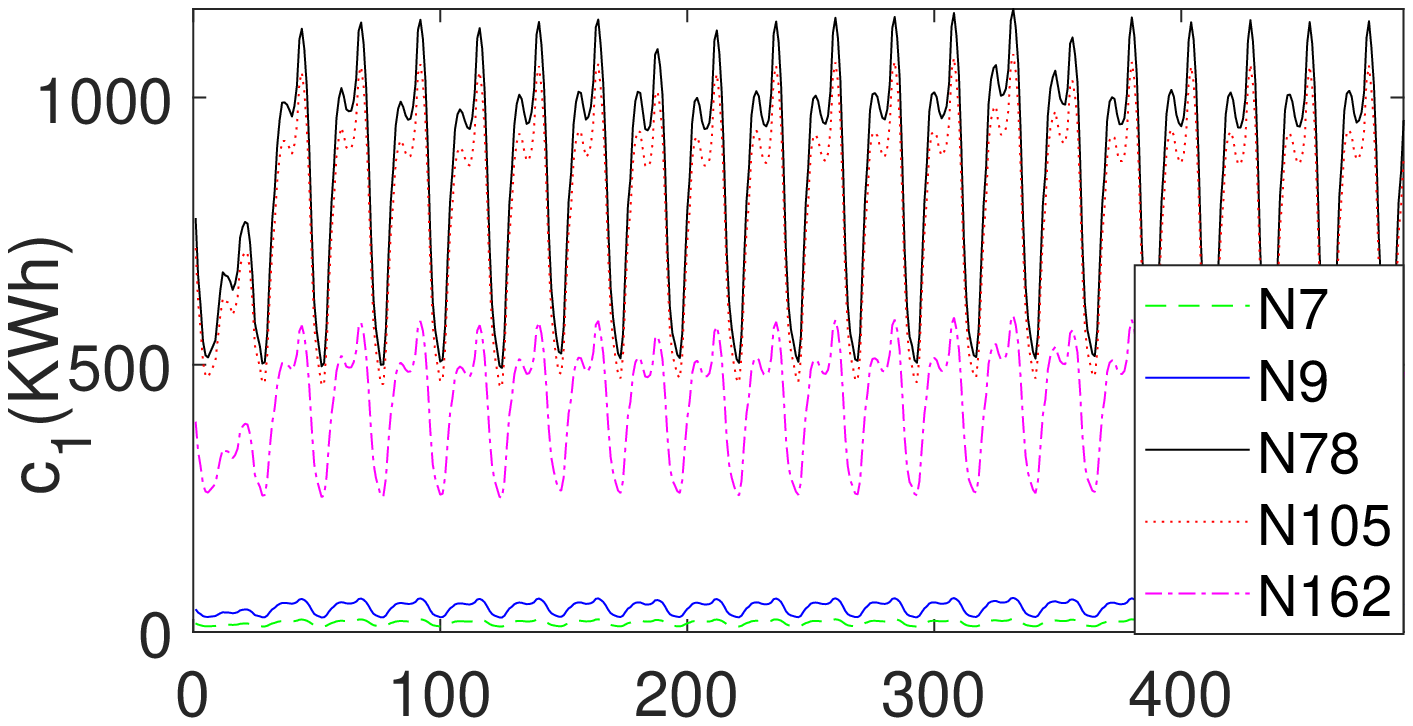} \hspace{-5mm}
\includegraphics[width=0.25\textwidth]{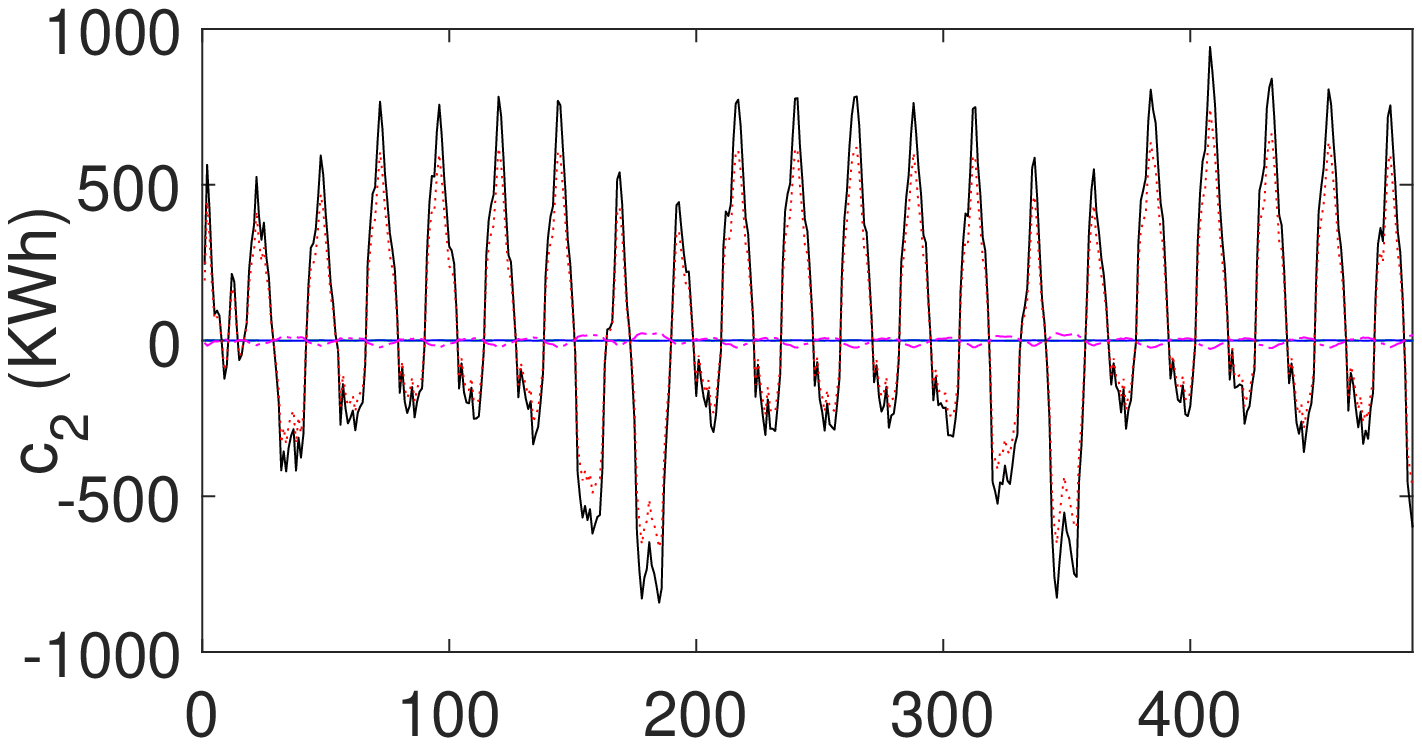} \\
\includegraphics[width=0.25\textwidth]{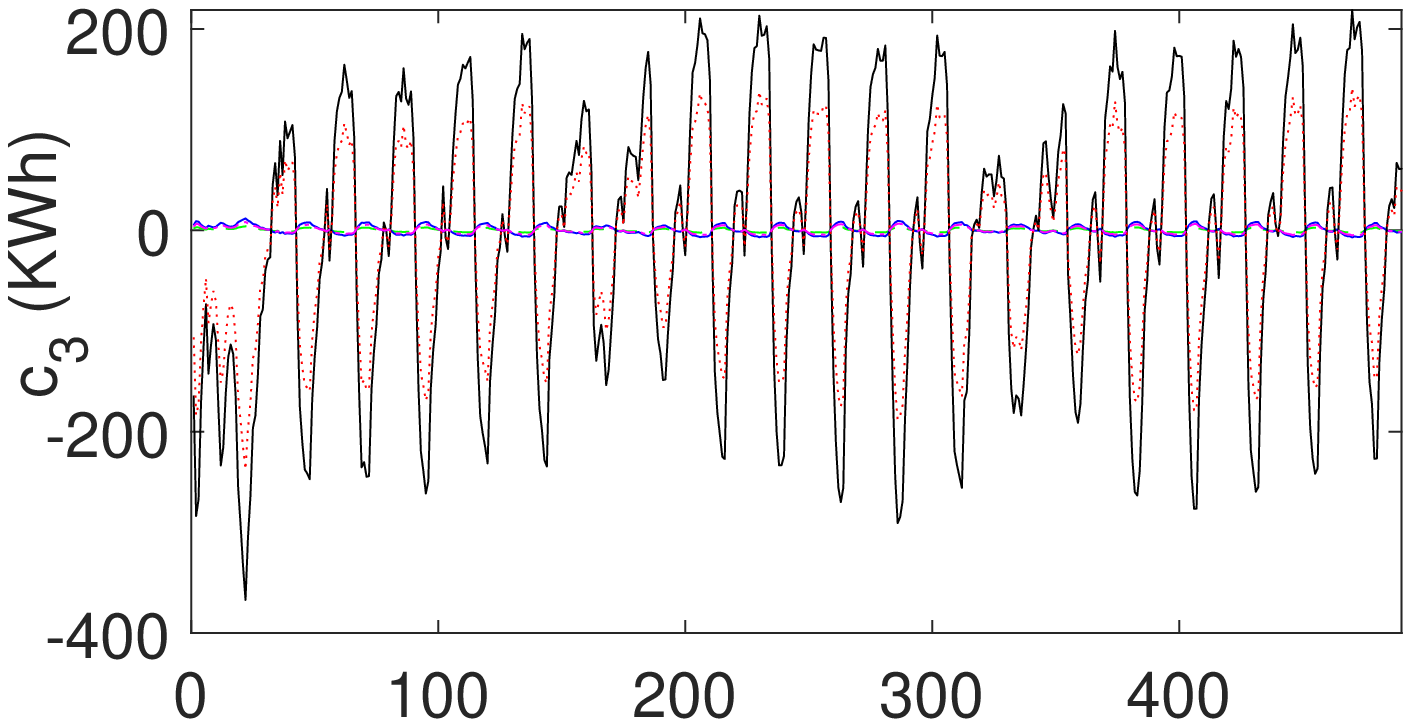} \hspace{-5mm}
\includegraphics[width=0.25\textwidth]{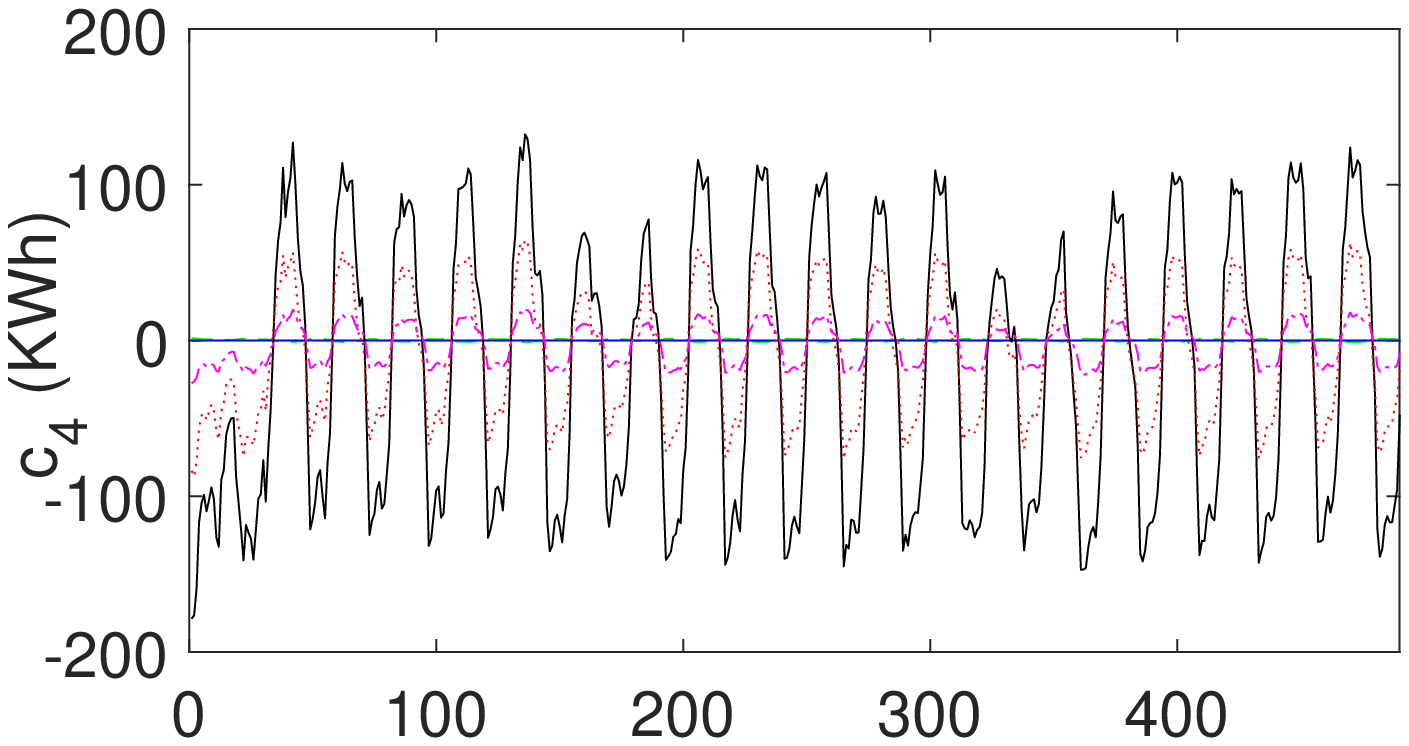} \\
\centering
\includegraphics[width=0.25\textwidth]{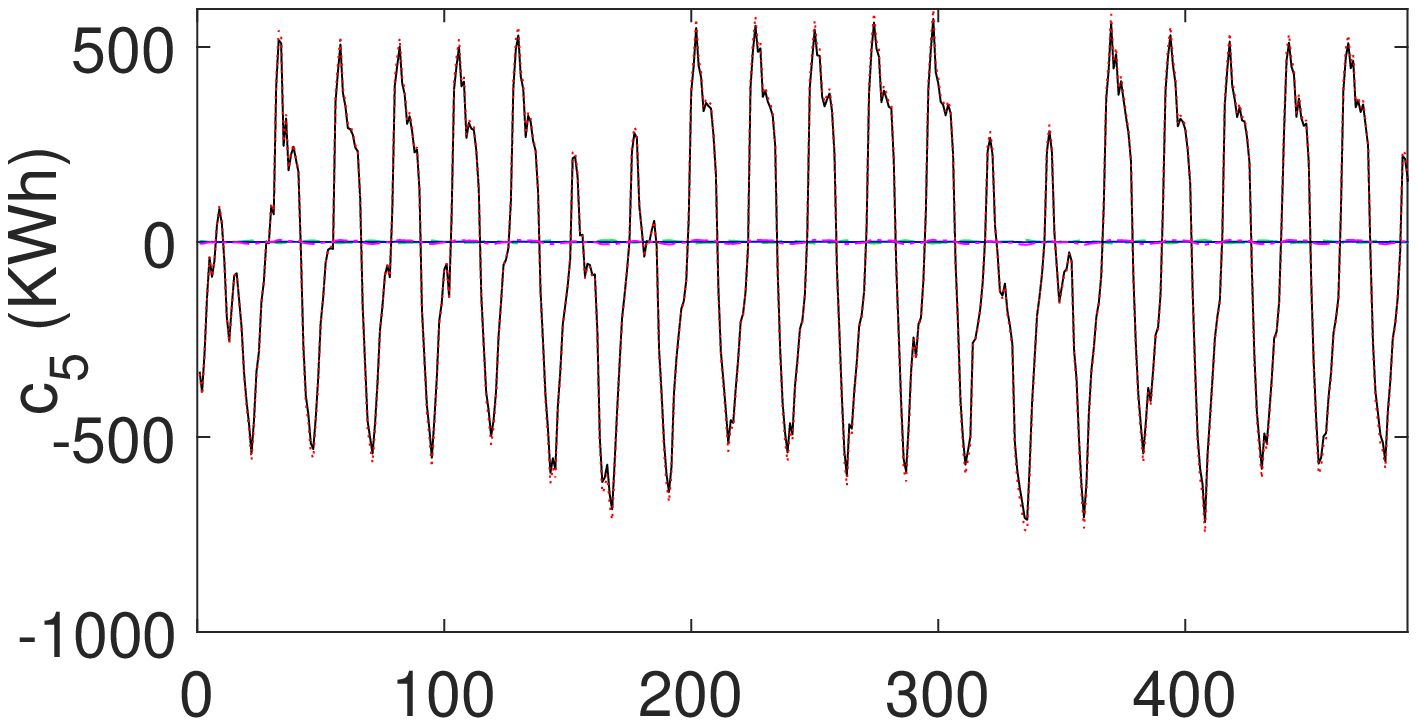} 
\caption{TGSD output from the electricity consumption data from Section V-C.}
\label{fig:tgsd_electric}
\end{figure}

The TVGMD method leverages the power of graph-theoretic tools to obtain inherent oscillatory signal components along with their network connectivity structures -- termed here as the \textit{graph modes}. A related approach, TGSD \cite{McNeil21}, also uses graph-related concepts to extract oscillatory signal components. Here we compare the results of the two methods on the electricity consumption data; the data set is explained in Section V-C and is plotted as the top left subplot of the Fig. \ref{fig:electricity2}. The oscillatory modes from the TVGMD and the TGSD methods are shown in Fig. \ref{fig:electricity2} and Fig. \ref{fig:tgsd_electric} respectively. The TGSD results shown here are based on its best-tuned parameters so that the obtained modes are physically meaningful. As noted earlier, each (graph) mode from the TVGMD method corresponds to the electricity consumption pattern at a specific time scale e.g., $c_5-c_1$ are associated with 6-hourly, 8-hourly, 12-hourly, daily and 1-weekly consumption patterns respectively. In contrast, the TGSD modes could not be associated to specific time scales and there is a clear evidence of multiple oscillations appearing in a single mode e.g., in $c_1$ for the $N78$ data. One can also observe similar information spanning multiple modes e.g., across $c_3-c_5$. For these reasons, it is difficult to extract meaningful information about users' electricity consumption patterns from the TGSD modes. The improved performance of TVGMD over TGSD could be attributed to the data-driven nature of the TVGMD i.e., no fixed dictionary atoms (basis functions) are used in the decomposition. More importantly, the TVGMD method extracts the multi-scale network connectivity patterns of the data concurrently with the oscillatory modes. TGSD and no other method, to our knowledge, is able to accomplish that. 

TVGMD method has been generally found to be reasonably robust to the changes in its parameters, except for the number of graph modes $K$. In Table \ref{tab:par}, suggested range of values for each parameter of TVGMD has been provided. The parameter $\alpha$ dictates the bandwidth of the graph modes: the lower its value, the higher the bandwidth of the modes. In our experiments, the range of values $\alpha=500-2000$ provided good results, depending on the whether the low-bandwidth or high-bandwidth graph modes are needed. Lower values of the parameter $\beta$ i.e., in the range of $0.1-1$, and $\gamma$, within the range of $0.5-1.0$, were found to produce optimal graph modes in our experiments. The nonzero value of the parameter $\tau$ enables the term involving Lagrangian multiplier in the proposed model, thus enforcing exact reconstruction of input signal via graph modes. The lower values of $\tau$ (including $\tau=0$), therefore, are useful in the case of noisy input signal where its exact reconstruction is not desirable. Finally, the choice of the number of graph modes $K$ was mainly dictated by the input data at hand and the performance of the algorithm depended heavily on the correct choice of $K$. For instance, in the case of synthetic graph signal \eqref{eq:syn} for which $K=4$ is known by design, choosing any other value of $K$ resulted in significant graph \textit{mode-mixing} (multiple oscillatory components in a single mode). This issue represents one of the limitations of the TVGMD method as the knowledge of correct value of $K$ may not be known a priori.

While the convergence of the TVGMD method has not been established and is not in the scope of this work, the TVGMD method converged in all the experiments shown in this work as well as in other simulations involving a wide range of signals.     


\begin{table}[!t]
	\renewcommand{\arraystretch}{1.3}
	\caption{User-defined parameter values within TVGMD}
	\label{tab:par}
	\centering
	\begin{tabular}{cccccc}
		\hline
		Parameters & $\alpha$ & $\beta$ &$\gamma$ &$\tau$ & $K$ \\
		\hline 
		TVGMD & 500-2000 & 0.1-1 & 0.5-1 &0-1 &input dependent \\
		\hline 
	\end{tabular}
\end{table}



\section{Conclusion}
In this paper, a new method for extracting the inherent modes of time-varying graph signals has been proposed. The extracted \textit{graph modes} are defined by the following useful pieces of information: i) the graph modes comprise inherent oscillatory signals present in the data at multiple nodes; ii) there is an alignment of information provided by the graph modes at each node (vertex); iii) the graph modes convey information about the underlying connectivity networks of the time-varying graph signals at multiple frequency scales. The method is the first-of-its-kind to provide the above information about the temporal graph signals concurrently by using a fully data-driven approach. By data-driven approach, it is meant that no predefined basis functions (either in the time- or graph-domains) have been used to obtain the \textit{graph modes}. 

At the heart of the proposed method is a robust variational optimization formulation that includes multiple requirements (or constraints) for the extraction of finite number of graph modes from an input graph signal. Those requirements relate to the properties of the graph modes both in the time-domain (e.g., limited bandwidth, complete signal reconstruction) and along the graph geodesics (e.g., smoothness of the graph modes along their respective adjacency matrices). We have proposed a method to solve the resulting optimization problem by using the alternative direction method of multipliers (ADMM) and the primal-dual technique. The proposed method outputs \textit{graph modes} that comprise the time-series associated with each node at multiple frequency scales and their corresponding network connectivity structure. 

To demonstrate the power and the potential of the proposed method in different domains, results from multiple experiments have been included in the paper. Particularly, the proposed algorithm has been shown to isolate alpha rhythms from the resting-state EEG signals and to correctly localize their sources in the occipital and parietal regions in the brain (See Fig. \ref{fig:eeg_graph}). Further, the proposed method was used to obtain the connectivity patterns of users in terms of their electricity consumption at different time scales (Fig. \ref{fig:electricity1}). Finally, the connectivity graphs of the flotation plant data set (Fig. \ref{fig:float_graphs}) could be used to define a grouping strategy of the plant cells, facilitating the fault diagnosis in flotation plants.      

Future work could consider extending this method to obtain time-varying adjacency (connectivity) matrices. Currently, the method obtains \textit{static} connectivity matrices based on the graph signal evolution for the entire signal duration. This is clearly a limitation of the proposed method since graph signals can exhibit non-stationarity in terms of time-varying connectivity patterns. Other avenues for future work may include: i) the development of computationally efficient approaches that target the extraction of a single graph mode at a time; ii) new adaptive approaches that automatically set the optimal number of graph modes $K$.  

\section{Appendix}
To solve the convex optimization problem given in \eqref{eq:primal-dual}, a primal-dual splitting approach is employed that jointly solves the primal and the associated dual optimization problems instead of focusing exclusively on either one. These algorithms are discussed in detail in \cite{Komodakis15}. Among these class of algorithms, the one used in this work is the \textit{forward-backward-forward} (FBF) based primal-dual algorithm that combines the gradient descent step(s) with the computation step(s) involving the proximity operator. The steps of that algorithm are listed in Algorithm 6 in \cite{Komodakis15}. In what follows, we list the steps to solve \eqref{eq:primal-dual} by using the FBF based primal-dual method.      

\begin{algorithm}
\caption{Primal-Dual Algorithm to Solve \eqref{eq:primal-dual}}\label{alg:fbf}
\text{Input: }{$z, \beta, \gamma, w_1^{(k)}, d_1\in\mathbb{R}_+^m, \delta, \epsilon, i\gets0 $}
\begin{algorithmic}
\Repeat
\State $i \gets i+1$
\State$y_{i+1} \gets w_i - \delta \times(2\gamma w_i + Q^Td_i
)$
\State$\overbar{y}_{i+1} \gets d_i + \delta\times(Qw_i)$
\State$p_{i+1} \gets \max(0,y_{i+1}-2\beta\delta z)$
\State$\overbar{p}_{i+1} \gets \frac{(\overbar{y}_{i+1}-\sqrt{\overbar{y}_{i+1}^2+4\delta})}{2}$
\State$q_{i+1} \gets p_{i+1} - \delta\times(2\gamma p_{i+1} +Q^T\overbar{p}_{i+1})$
\State$\overbar{q}_{i+1} \gets \overbar{p}_{i+1}+\delta\times(Qp_{i+1})$
\State$(w_{i+1},d_{i+1})=(w_i-y_{i+1}+q_{i+1},w_i-\overbar{y}_{i+1}+\overbar{q}_{i+1})$
\Until{Convergence: $\frac{\Vert w_{i+1}-w_{i}\Vert_2}{\Vert w_{i}\Vert_2}<\epsilon \And \frac{\Vert d_{i+1}-d_{i}\Vert_2}{\Vert d_{i}\Vert_2}<\epsilon$}
\end{algorithmic}
\end{algorithm}

\bibliographystyle{IEEEtran}
\bibliography{IEEEabrv,ref.bib}
\end{document}